\documentclass[10pt,letterpaper]{article}
\usepackage[margin=1in]{geometry}
\usepackage[utf8]{inputenc}
\usepackage[T1]{fontenc}
\usepackage{lmodern}
\usepackage{microtype}
\usepackage{graphicx}
\usepackage{amsmath,amssymb,amsfonts}
\usepackage{amsthm}
\usepackage{mathrsfs}
\usepackage[title]{appendix}

\usepackage{textcomp}
\usepackage{booktabs}
\usepackage{multirow}
\usepackage{array}
\usepackage{makecell}
\usepackage{arydshln}
\usepackage{seqsplit}
\usepackage{algorithm}
\usepackage{algorithmicx}
\usepackage{algpseudocode}
\usepackage{listings}
\usepackage{caption}
\usepackage[numbers,sort&compress]{natbib}
\usepackage{hyperref}
\hypersetup{colorlinks=true,linkcolor=black,citecolor=black,urlcolor=black}
\usepackage{xcolor}
\usepackage{aiayn-title}

\theoremstyle{plain}

\theoremstyle{definition}

\begin{document}

\title{Adaptable Method for Crystal Design \\across Diverse Constraints and Objectives \\with Pretrained Property Predictors}

\author{
Akihiro Fujii\\
The University of Tokyo\\
\texttt{akihiro.fujii@cello.t.u-tokyo.ac.jp}
\And
Yoshitaka Ushiku\\
OMRON SINIC X Corp.\\
\texttt{contact@yoshitakaushiku.net}
\AND
Anh Khoa Augustin Lu\\
The University of Tokyo  \\ National Institute for Materials Science (NIMS) \\
\texttt{lu.augustin@cello.t.u-tokyo.ac.jp}
\And
Koji Shimizu\\
National Institute of Advanced Industrial Science and Technology (AIST)\\
\texttt{koji.shimizu@aist.go.jp}
\AND
Satoshi Watanabe\\
The University of Tokyo\\
\texttt{watanabe@cello.t.u-tokyo.ac.jp}
}

%\vspace*{0.35in}
%\begin{flushleft}
%{\Large\textbf{Adaptable method for crystal design across diverse constraints and objectives with pretrained property predictors}}
%\\[1.0em]
%Akihiro Fujii\textsuperscript{1,*}, Yoshitaka Ushiku\textsuperscript{2}, Anh Khoa Augustin Lu\textsuperscript{1,3}, Koji Shimizu\textsuperscript{4}, and Satoshi Watanabe\textsuperscript{1}
%\\[1.0em]
%\textbf{1} Department of Materials Engineering, The University of Tokyo, Faculty of Engineering Bldg. IV, 7-3-1, Hongo, Bunkyo-ku, Tokyo 113-8656, Japan\\
%\textbf{2} OMRON SINIC X Corp., Nagase Hongo Building 3F, 5-24-5 Hongo, Bunkyo-ku, Tokyo 113-0033, Japan\\
%\textbf{3} National Institute for Materials Science (NIMS), 1-1 Namiki, Tsukuba, Ibaraki 305-0044, Japan\\
%\textbf{4} National Institute of Advanced Industrial Science and Technology (AIST), 1-1-1 Umezono, Tsukuba, Ibaraki 305-8568, Japan\\%[1.0em]
%* Corresponding author: \href{mailto:akihiro.fujii@cello.t.u-tokyo.ac.jp}{akihiro.fujii@cello.t.u-tokyo.ac.jp}
%\end{flushleft}
\maketitle

\begin{abstract}
Advanced crystal design can accelerate materials discovery across applications from photovoltaics to spintronics. Practical design must satisfy multiple properties and physical constraints, yet existing machine-learning-based approaches to such design often depend on large datasets, retraining, or task-specific generators. Here, we show that direct predictor-guided gradient optimization enables data-efficient, constraint-rich crystal design by combining off-the-shelf predictors with site-wise element masks, template initialization, and task-specific losses. In perovskites, it outperformed generative and Bayesian baselines under three targets—band gap, formation energy, and tolerance factor—and two hard constraints. DFT assessment further showed band-gap targeting competitive with a leading generative model despite using predictors trained on roughly one-tenth of the data. By flexibly combining pretrained predictors with application-oriented masks and custom losses, the same framework supported half-metal design. Such modularity could help researchers and engineers translate diverse application requirements directly into optimized candidate crystals with minimal computational cost.
\end{abstract}

\vspace{0.7em}

\begin{figure}[t]
\begin{center}
\includegraphics[width=0.85\linewidth]{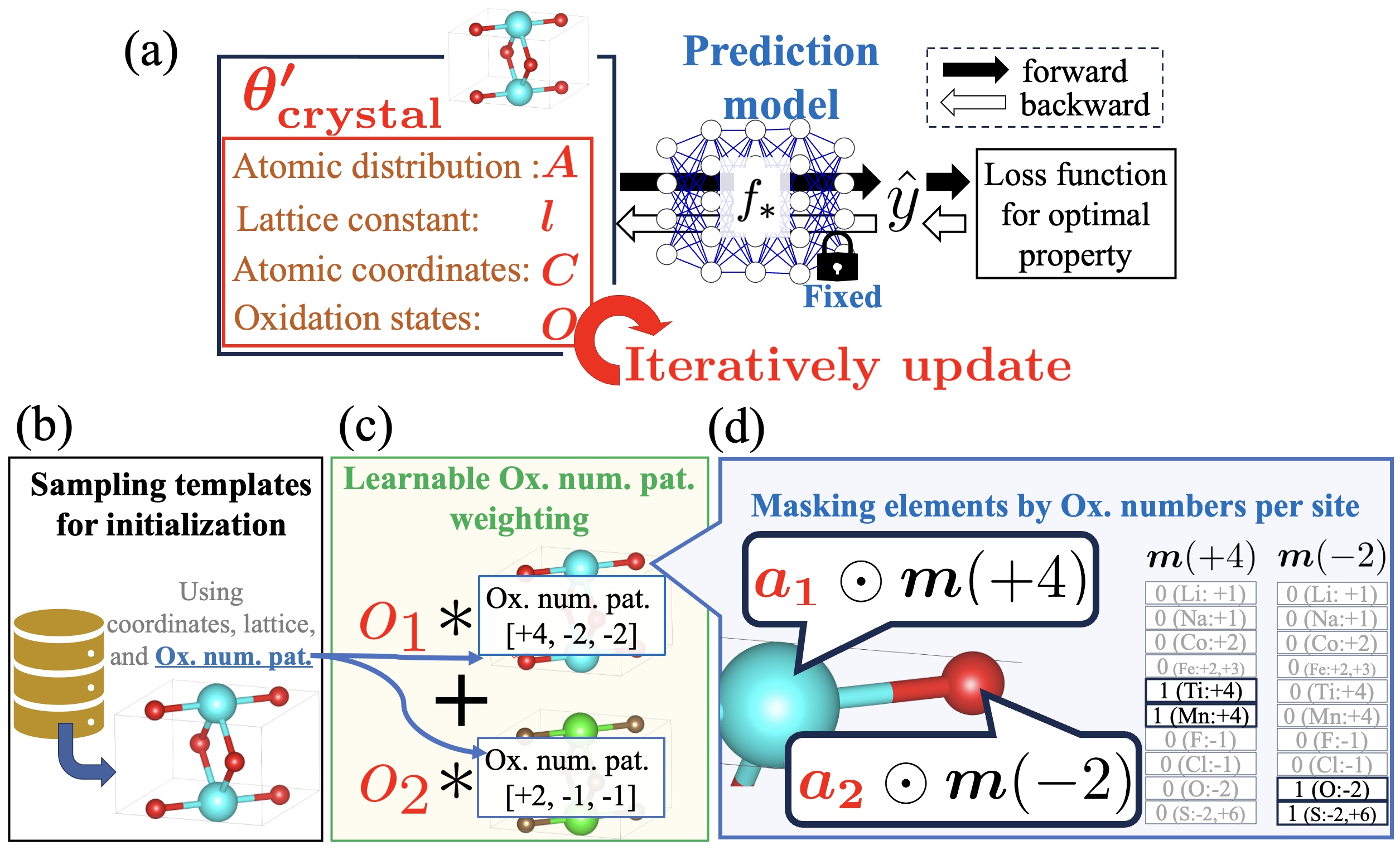}
\end{center}
\caption{Overview of the SMOACS framework. (a) SMOACS directly optimizes crystal structures using gradients from pretrained property predictors. (b) Initial lattice, coordinates, and oxidation number patterns (Ox. num. pat.) are extracted from template structures. (c) Atomic distributions are computed as weighted sums of two oxidation patterns, $[+4,-2,-2]$ and $[+2,-1,-1]$, using learnable weights $o_1$ and $o_2$. (d) SMOACS enforces charge neutrality by masking elements based on oxidation numbers $\boldsymbol{m}(+4)$ and $\boldsymbol{m}(-2)$. Each mask value is 1 if the element with the given oxidation number is allowed, and 0 otherwise. Here, $\odot$ denotes element-wise multiplication.}
\label{fgr__overview}
\end{figure}

\section{Introduction}\label{sec1}

Designing crystalline materials that meet multiple property targets under constraints remains a major challenge in materials science. Applications ranging from perovskite photovoltaics to spintronics impose distinct requirements, including suitable band gaps, stability, and spin-resolved transport. Additionally, practical deployment can require avoiding toxic elements or elements with high supply-chain risk. Exhaustive search based on density functional theory (DFT) calculations~\cite{nishijima2014accelerated} or machine-learning predictors trained on DFT-generated data~\cite{merchant2023scaling} can screen candidates before synthesis, but the vast combinatorial design space demands more efficient exploration. Furthermore, modeling dopants and substitutions often requires larger cells, further expanding the design space. Thus, a useful framework should efficiently explore the design space while optimizing multiple properties, scaling to larger structures, and adapting to changing specifications without task-specific model retraining.

Recent progress has been driven by black-box optimization~\cite{herbol2018efficient, karpovich2024deep} and deep generative modeling~\cite{REN2022314, xiecrystal, zeni2023mattergen}, moving beyond exhaustive search. Among black-box methods, Bayesian optimization can offer strong sample efficiency for expensive objectives and remains an important baseline for materials design~\cite{brochu2010tutorial, boyar2024crystal, zhai2024benchmarking, khatamsaz2023physics, ozaki2020automated}, while reinforcement-learning-based approaches have also recently shown promise~\cite{govindarajan2024learning}. Deep generative models, including variational, flow- and diffusion-based approaches, have become a major route by learning distributions over crystal structures, sometimes with predictor guidance or conditional generation~\cite{REN2022314,xiecrystal, zeni2023mattergen, yang2024scalable, zhu2024wycryst, luo2025crystalflow}. 

Yet constraint-rich materials design exposes limitations in both families. Black-box methods such as Bayesian optimization and reinforcement learning become increasingly difficult as the search space grows and can suffer from inefficient early exploration, i.e., the \textit{cold-start problem}~\cite{barto2003recent}. Generative approaches are suited to high-dimensional crystal spaces, but high-quality design typically requires large and diverse training data. Furthermore, adapting them to new specifications often requires costly retraining~\cite{zeni2023mattergen}.

These considerations motivate revisiting direct gradient-based optimization in crystal-structure space, an approach used in shape design, image editing, and compositional design~\cite{allen2022physical, xia2022gan, fujii2025straightforward}. Successful large-scale optimization~\cite{aage2017giga} suggests that gradients can guide high-dimensional crystal-structure searches. Pretrained predictors can provide surrogate landscapes that mitigate cold starts, and new objectives can be introduced by reusing off-the-shelf predictors~\cite{Kamal2021alignn, taniaicrystalformer}. Hard constraints can be imposed through the optimization domain and initialization scheme, whereas soft constraints can be encoded in the loss. Large structural templates can further support the initialization of doped or substituted structures.
%These con∂çsiderations motivate revisiting direct gradient-based optimization in crystal-structure space, an approach effective in shape design, image editing, and compositional design~\cite{allen2022physical,  xia2022gan, bordiga2024automated, fujii2025straightforward}. First, for materials design, where the search space is inherently high-dimensional, this strategy is appealing because the success of optimization in large-scale design~\cite{aage2017giga} suggests that exploiting gradients may also be effective for high-dimensional crystal-structure optimization. Second, reusing pretrained predictors provides informative surrogate landscapes, thereby helping mitigate the cold-start problem. Third, it offers low-cost adaptability: hard constraints can be imposed by redefining the optimization domain and initialization scheme, while soft constraints can be handled through loss functions that reuse and combine off-the-shelf property predictors~\cite{Kamal2021alignn, taniaicrystalformer}. Finally,large structural templates provide initial configurations for optimizing doped or substituted structures.

For crystalline materials, however, this idea has remained underdeveloped. This may be because naive gradient descent is vulnerable to rugged objective landscapes and poor local minima. In addition, many constraints in materials design---charge neutrality, project-defined element-exclusion constraints, and preservation of a prescribed structural motif---are inherently non-differentiable. 

Here we address these obstacles with Simultaneous Multi-property Optimization using Adaptive Crystal Synthesizer (SMOACS), a predictor-guided framework for constraint-rich crystal design (Figure~\ref{fgr__overview}). Building on atomic-distribution representations~\cite{xiecrystal, zeni2023mattergen} that relax discrete elemental identities into differentiable site-wise distributions, SMOACS updates crystal variables using gradients from pretrained predictors and introduces two main design choices---site-wise element masks and template-based initialization---to address non-differentiable constraints and poor local minima. The contribution of SMOACS is not the atomic-distribution relaxation or masking of such distributions~\cite{fujii2025straightforward}, but their integration with learnable oxidation-pattern weights, template-based initialization, custom losses, and off-the-shelf predictors for direct, constraint-rich crystal optimization. Within this framework, site-wise element masks restrict the atomic distributions to admissible elements, including oxidation-state requirements for charge neutrality and exclusions based on toxicity or supply-chain risk, while learnable oxidation-pattern weights allow multiple charge-neutral oxidation-state patterns to be considered simultaneously. Template-based initialization exploits the tendency of crystal structures to cluster in configuration space, as exemplified by rutile- and perovskite-type structures, thereby keeping the search in plausible regions, reducing convergence to poor local minima, and preserving the target motif within restricted coordinate ranges. SMOACS thus enables high-dimensional, constraint-rich design by combining custom differentiable loss functions, application-oriented masks, and off-the-shelf predictors without retraining.

%Here we address these obstacles with Simultaneous Multi-property Optimization using Adaptive Crystal Synthesizer (SMOACS), a framework that makes direct predictor-guided crystal optimization practical for constraint-rich materials design (Figure~\ref{fgr__overview}). SMOACS updates crystal variables using gradients from pretrained predictors and introduces two main design choices---site-wise element masks and template-based initialization---to address non-differentiable constraints and poor local minima. First, site-wise element masks impose site-specific admissibility constraints. These masks can enforce oxidation-state requirements for charge neutrality and exclude toxic elements or elements with high supply-chain risk. Second, template-based initialization starts from known structural motifs, keeping the search in plausible regions and avoiding poor local minima. This reflects the observation that many crystal structures occupy similar regions in configuration space, rather than being randomly distributed, as exemplified by rutile- and perovskite-type structures. Together, template-based initialization and a restricted optimization range help preserve the prescribed structural motif. We also use mechanisms to escape local minima, including simulated annealing and mutation. Overall, SMOACS enables high-dimensional, constraint-rich materials design by flexibly combining custom differentiable losses, application-oriented masks, and off-the-shelf predictors without retraining.

We first show that site-wise element masks and template-based initialization improve gradient-based crystal search. We then evaluate SMOACS in two constraint-rich design tasks and a separate data-efficiency comparison with MatterGen. In constrained perovskite design, SMOACS outperforms generative and black-box baselines under coupled property targets and explicit chemical and structural constraints, and scales to 135-atom perovskites. Compared with MatterGen~\cite{zeni2023mattergen}, it remains competitive despite relying on predictors trained on roughly one-tenth of the data. Finally, a constrained half-metal case study shows that application-oriented masks and custom loss functions can configure SMOACS for realistic design tasks. These results position direct gradient-based crystal optimization as a practical route for constraint-rich materials design when data and computational resources are limited.
%We first show that site-wise element masks and template-based initialization improve gradient-based crystal search. We then examine SMOACS in three settings reflecting constraint-rich materials design. Using perovskite design as a practical, constraint-rich test case, SMOACS achieves higher success rates than generative and black-box baselines under coupled property targets and explicit chemical and structural constraints, and also succeeds in optimizing 135-atom perovskite structures. Compared with MatterGen~\cite{zeni2023mattergen}, a leading generative model trained on a much larger dataset, SMOACS remains competitive despite using predictors trained on substantially less data. Finally, a half-metal design case study with application-oriented masks and custom losses demonstrates that SMOACS can be configured for realistic design tasks. Together, these results position direct gradient-based crystal optimization as a practical route for constraint-rich materials design when data and computational resources are limited.

%
%
% -------------------------------------------------------
% -------------------------------------------------------
% -------------------------------------------------------
%
%

\section{Results} \label{sectin___experiments}

\begin{table}[t]
\centering
\caption{Results for simultaneous optimization of two target properties and one constraint: band gap, formation energy, and charge neutrality. ``Rand.'' and ``Templ.'' in ``Struct. Init.'' denote random and template-based initialization (from the MEGNet dataset~\cite{chen2019graph}), and ``Ox. Restr.'' indicates whether an oxidation-number mask is applied. ``SA'' and ``Mut.'' indicate simulated annealing and mutation, respectively. ``Success rate'' is the joint probability of satisfying (i) $E_g \in 2.5 \pm 0.2$ eV, (ii) $E_f < -0.5$ eV/atom, and (iii) charge neutrality (``Neu.''). ``Uniq.'' is the fraction of unique element combinations. ``Novel in SC'' indicates the fraction of successful samples absent from the MEGNet dataset. Averaged over four runs with 128 samples each. See Supplementary Information (SI) section~\ref{appendix__details_of_init_test} for the full results.}
\label{table__summary_initialization_mutation_analysis}
\begin{tabular}{wl{2.8cm}wl{1.1cm}wl{1.4cm}wc{1.3cm}wc{0.95cm}wc{0.95cm}wc{0.95cm}wc{0.85cm}wc{1.05cm}}
\toprule
Opt. Method & \makecell[l]{Struct.\\Init.} & Ox. Restr. & \makecell{Success\\Rate} & (i) $E_g$ & (ii) $E_f$ & (iii) Neu. & Uniq. & \makecell{Novel\\in SC} \\
\midrule
Grad. & Rand. & None & $0.00$ & 0.30 & 0.06 & 0.01 & N/A & N/A \\
Grad. + SA &  Rand. & None &$0.00$ & 0.47 & 0.12 & 0.01 & N/A & N/A \\
Grad. & Templ. & None  & $0.07$ & 0.65 & 0.83 & 0.11 & 0.97 & 0.93 \\
Grad. & Templ. &Ox. Mask &  $0.42$ & 0.58 & 0.62 & $\underline{\mathbf{1.00}}$ & 1.00 & 0.96 \\
Grad. + SA & Templ. &Ox. Mask & $0.47$ & 0.52 & 0.75 & $\underline{\mathbf{1.00}}$ & 0.99 & 0.88 \\
Grad. + Mut. & Templ. &Ox. Mask & $0.64$ & 0.67 & 0.93 & $\underline{\mathbf{1.00}}$ & 0.99 & 0.92 \\
Grad. + Mut. + SA & Templ. &Ox. Mask & $\underline{\mathbf{0.67}}$ & $\underline{\mathbf{0.70}}$ & $\underline{\mathbf{0.94}}$ & $\underline{\mathbf{1.00}}$ & $\underline{\mathbf{1.00}}$ & $\underline{\mathbf{0.97}}$ \\
\bottomrule
\end{tabular}
\end{table}

We evaluate SMOACS as follows: an ablation of template initialization, site-wise element masks and strategies for escaping poor local minima; constrained perovskite optimization; comparison with MatterGen~\cite{zeni2023mattergen} using predictors trained on much less data; and a constrained half-metal candidate search as a practical demonstration of SMOACS.

\subsection{Impact of template-based initialization and element masks on multi-targeted materials design}
\label{section__init_experiment}

We first assess the role of the proposed optimization ingredients using a simpler multi-target task. Using a Crystalformer~\cite{taniaicrystalformer} predictor trained on the MEGNet dataset~\cite{chen2019graph}, we searched for crystals that simultaneously satisfied a target band gap of $2.5 \pm 0.2$ eV, a formation energy below $-0.5$ eV/atom, and charge neutrality over 200 optimization steps. Charge neutrality was defined by a zero sum of oxidation numbers over all sites (SI section~\ref{appendix__electrical_neutrality}). This task also directly tests the two components introduced in Figure~\ref{fgr__overview}: template-based initialization to avoid poor local minima, and site-wise element masks based on oxidation numbers to enforce charge neutrality, a non-differentiable constraint. We also added mutation and simulated annealing (SA) to improve exploration. For all methods, we evaluated whether generated structures satisfied the target criteria using the same predictor model employed during optimization.

Table~\ref{table__summary_initialization_mutation_analysis} shows that each ingredient addresses a distinct failure mode. Random structural initialization was ineffective: plain gradient descent gave a success rate of $0.00$, and SA alone did not rescue the search. Switching only the initialization from random to template-based markedly improved the band-gap and formation-energy success rates, leading to an overall success rate of $0.07$, indicating that the starting point is critical in crystal-space optimization. Adding the oxidation-number-based site-wise mask then drastically increased the success rate to $0.42$ while bringing charge neutrality to $1.00$. Mutation and SA brought further gains, ultimately increasing the success rate to $0.67$. Successful candidates remained highly unique, and many were novel with respect to the MEGNet training set.

Overall, success is driven primarily by template-based initialization and site-wise masking, with mutation providing a substantial further gain and simulated annealing adding a smaller additional improvement.

\subsection{Multi-property design of perovskite structures}
\label{sec__perov_experiments}

We next examined constrained optimization of perovskites as the first core benchmark. In its prototypical form, a perovskite adopts a cubic \(\mathrm{ABX_3}\) framework, with A-site cations at the body center, B-site cations at the cube corners, and X-site anions at the edge centers. We optimized three properties---band gap, formation energy, and the Goldschmidt tolerance factor---under two hard design requirements: charge neutrality and preservation of the perovskite structure. The Goldschmidt tolerance factor is a widely used indicator of structural stability in \(\mathrm{ABX_3}\) perovskites and is calculated from the ionic radii of the A, B, and X sites. A candidate therefore counted as successful only when it simultaneously (i) matched the requested band-gap window, (ii) had sufficiently low formation energy, (iii) fell in the accepted tolerance-factor range, (iv) remained charge-neutral, and (v) preserved the geometric features of a perovskite. 

In this experiment, we performed optimization at two structural scales: a unit-cell perovskite structure and a 135-site cell obtained by \(3 \times 3 \times 3\) expansion of the perovskite unit cell. Evaluating the \(3 \times 3 \times 3\) cell further allowed us to test whether the method could propose structures involving elemental substitution or dopant incorporation. Since each site was optimized over three fractional coordinates and approximately 100 candidate elemental species, the effective search dimensionality was roughly 500 for the unit cell and 13{,}500 for the \(3 \times 3 \times 3\) cell. In the \(3 \times 3 \times 3\)-cell experiments, charge neutrality was not included among the success criteria because exhaustive charge-neutrality enumeration is combinatorially intractable.

In this benchmark, SMOACS enforced design requirements directly in structure space: perovskite fidelity through the choice of initialization and the admissible optimization range, and charge neutrality through site-wise element masks. To assess the applicability of SMOACS, we conducted these experiments using both ALIGNN~\cite{Kamal2021alignn}, a graph-neural-network-based model, and Crystalformer~\cite{taniaicrystalformer}, a Transformer-based model. The objective then combined predictor-based band-gap and formation-energy terms from ALIGNN or Crystalformer with a Goldschmidt tolerance-factor term computed from the atomic distributions. The coordinate-deviation tolerance from the ideal perovskite positions was set to $\epsilon = 0.15$ for the unit cell and $\epsilon = 0.05$ for the $3\times3\times3$-size cell, based on a reported distorted perovskite structure. Ionic radii used in the tolerance-factor calculation were taken from pymatgen~\cite{ONG2013314}, and the tolerance-factor success window was set to $0.8 \leq t \leq 1.0$ based on representative perovskite structures. See SI section~\ref{appendix__details_of_perov_crystal_bg_opt} for details. As reference methods, we used the generative baselines FTCP~\cite{REN2022314} and CDVAE~\cite{xiecrystal} together with the tree-structured Parzen estimator (TPE)~\cite{ozaki2020automated} as a black-box optimizer.

\begin{table}
\centering
\caption{Experiments optimizing perovskite structures. For unit-cell experiments, ``SC rate'' refers to the proportion of samples that simultaneously satisfy five criteria: (i) the band gap falls within the target range, (ii) formation energy is below \(-0.5\) eV/atom, (iii) the tolerance factor \(t \in [0.8, 1.0]\), (iv) the structure is charge-neutral, and (v) the structure approximates a perovskite. For \(3 \times 3 \times 3\)-cell experiments, charge neutrality was not included among the success criteria because exhaustive charge-neutrality enumeration is combinatorially intractable. Criterion (v) is fulfilled when both of the following hold: (a) internal coordinates deviate by less than $\pm \epsilon$ from those of a typical perovskite, and (b) crystal-axis angles are between $85^\circ$ and $95^\circ$. ``Neu.'' is reported where evaluated, but excluded from the 3×3×3 SC rate. Averaged over four runs with 128 samples each. See Table~\ref{table__perovskite_eval_all} and \ref{table__3x3x3_supercell_all} for augmented results.}
\label{table__perovskite_eval}
\begin{tabular}{llcccccc}
\toprule
\makecell{Target $E_g$ \\ (cell size)} & Method & \makecell{SC\\rate} & \makecell{(i)\\$E_g$} & \makecell{(ii)\\$E_f$} & \makecell{(iii)\\$t$} & \makecell{(iv)\\Neu.} & \makecell{(v)\\Prv.} \\
\midrule

\multirow{5}{*}{\makecell{$0.5 \pm 0.2$ \\ eV \\ (unit cell)}} 
& FTCP    & $0.00$ & $0.00$ & $1.00$ & $0.23$ & $0.99$ & $0.23$ \\
\cmidrule(lr){2-8}
& CDVAE   & $0.00$ & $0.21$ & $0.42$ & $0.00$ & $0.38$ & $0.00$ \\
\cmidrule(lr){2-8}
& TPE     & $0.09$ & $1.00$ & $0.46$ & $0.30$ & $0.95$ & $1.00$ \\
\cmidrule(lr){2-8}
& S(ALI)  & $0.20$ & $0.71$ & $0.46$ & $0.44$ & $1.00$ & $1.00$ \\
\cmidrule(lr){2-8}
& S(Cry)  & $\underline{\mathbf{0.23}}$ & $0.62$ & $0.86$ & $0.37$ & $1.00$ & $1.00$ \\
\midrule

\multirow{5}{*}{\makecell{$4.0 \pm 0.2$ \\ eV \\ (unit cell)}} 
& FTCP    & $0.00$ & $0.00$ & $1.00$ & $0.22$ & $0.99$ & $0.22$ \\
\cmidrule(lr){2-8}
& CDVAE   & $0.00$ & $0.03$ & $0.42$ & $0.00$ & $0.38$ & $0.00$ \\
\cmidrule(lr){2-8}
& TPE     & $0.04$ & $0.96$ & $0.43$ & $0.20$ & $0.40$ & $1.00$ \\
\cmidrule(lr){2-8}
& S(ALI)  & $\underline{\mathbf{0.36}}$ & $0.60$ & $0.99$ & $0.62$ & $1.00$ & $1.00$ \\
\cmidrule(lr){2-8}
& S(Cry)  & $0.29$ & $0.62$ & $0.98$ & $0.46$ & $1.00$ & $1.00$ \\
\midrule

\multirow{5}{*}{\makecell{$0.5 \pm 0.2$ \\ eV \\ ($3\times3\times3$)}} 
& FTCP    & $0.00$ & $0.84$ & $1.00$ & $0.00$ & N/A & $0.00$ \\
\cmidrule(lr){2-8}
& CDVAE   & $0.00$ & $0.06$ & $0.20$ & $0.00$ & N/A & $0.00$ \\
\cmidrule(lr){2-8}
& TPE & $0.00$ & $1.00$ & $0.00$ & $0.99$ & N/A & $1.00$ \\
\cmidrule(lr){2-8}
& S(ALI)  & $\underline{\mathbf{0.36}}$ & $0.92$ & $1.00$ & $0.40$ & $1.00$ & $1.00$ \\
\cmidrule(lr){2-8}
& S(Cry)  & $0.05$ & $0.93$ & $0.10$ & $0.39$ & $1.00$ & $1.00$ \\
\midrule

\multirow{5}{*}{\makecell{$4.0 \pm 0.2$ \\ eV \\ ($3\times3\times3$)}} 
& FTCP    & $0.00$ & $0.00$ & $1.00$ & $0.00$ & N/A & $0.00$ \\
\cmidrule(lr){2-8}
& CDVAE   & $0.00$ & $0.00$ & $0.24$ & $0.00$ & N/A & $0.00$ \\
\cmidrule(lr){2-8}
& TPE & $0.00$ & $0.00$ & $0.00$ & $0.22$ & N/A & $1.00$ \\
\cmidrule(lr){2-8}
& S(ALI)  & $\underline{\mathbf{0.42}}$ & $0.75$ & $1.00$ & $0.47$ & $1.00$ & $1.00$ \\
\cmidrule(lr){2-8}
& S(Cry)  & $0.03$ & $0.17$ & $0.11$ & $0.44$ & $1.00$ & $1.00$ \\

\bottomrule
\end{tabular}
\end{table}

Table~\ref{table__perovskite_eval} shows that this joint criterion sharply separates SMOACS from the baselines. Here, S(ALI) and S(Cry) denote SMOACS driven by ALIGNN and Crystalformer, respectively. FTCP~\cite{REN2022314} and CDVAE~\cite{xiecrystal} remained at $0.00$ success rate for both target band gaps, despite FTCP often finding low-formation-energy candidates and CDVAE using predictor guidance in latent space. Notably, CDVAE achieves non-zero success in the simpler single-property benchmark (SI section~\ref{appendix__section__rare_experiment}), and prior demonstrations of generative models, including FTCP, CDVAE, and MatterGen, have involved at most three jointly optimized properties and constraints~\cite{REN2022314, xiecrystal, zeni2023mattergen}. Thus, the zero joint success rate here reflects the difficulty of the multi-objective, constraint-rich task. TPE matched the band-gap window frequently and preserved the predefined perovskite geometry, but its joint success was only modest in the unit-cell setting and dropped to zero for the $3\times3\times3$ setting. By contrast, SMOACS retained non-zero success across all four settings, and the best SMOACS variant consistently outperformed every baseline. Further details are provided in SI sections~\ref{appendix__details_of_perov_crystal_bg_opt} and~\ref{appendix__details_of_large_system}.

The failure modes of the baselines help explain this difference. FTCP and CDVAE search in latent space, which makes it difficult to preserve charge balance and perovskite geometry simultaneously (SI section~\ref{section__cdvae_analysis}). TPE searches within a predefined admissible region that enforces the structural constraint, but, as a black-box optimizer, it is prone to cold-start issues and inefficient exploration in high-dimensional spaces. By contrast, SMOACS uses predictor gradients to guide efficient search in high-dimensional structure space, while enforcing the admissible perovskite design space through templates, site-wise masks, and coordinate restrictions. Optimization results using predictors trained on two different datasets are provided in SI section~\ref{section__cross_dataset}.

\subsection{Comparison with a Leading Generative Model}
\label{sec__comparison_with_mattergen}

We next moved from predictor-level optimization to DFT-assessed candidate quality by comparing SMOACS with MatterGen~\cite{zeni2023mattergen}, a leading generative model for inorganic crystal design. The core of this comparison is the data asymmetry. For MatterGen, we used the publicly released model pretrained on 608{,}000 structures~\cite{zeni2023mattergen}, whereas SMOACS used ALIGNN band-gap and formation-energy predictors, each trained on 60{,}000 structures from the MEGNet dataset~\cite{chen2019graph}. We therefore treat this experiment as a test of whether direct, constraint-aware optimization can remain competitive in a markedly lower-data regime. For both methods, candidates were targeted toward five band-gap windows---$2.5$, $3.0$, $3.5$, $4.0$, and $4.5$ eV---and their band gaps were then re-evaluated by DFT. For the comparison with MatterGen under space-group conditioning, see SI section~\ref{appendix___spg_conditioned_generation}.

We adopt the Stability--Uniqueness--Novelty (S.U.N.) criterion~\cite{zeni2023mattergen}. Stability is defined by the energy distance from the convex hull ($E_\mathrm{hull}$), computed from MatterSim energies ($E^\mathrm{MatterSim}_\mathrm{hull}$) with the Materials Project~\cite{10.1063/1.4812323} as the reference set. This proxy was reliable, as the MatterSim-derived \(E_{\mathrm{hull}}\) values differed from the DFT-derived \(E_{\mathrm{hull}}\) values by a mean absolute error of \(0.017~\mathrm{eV/atom}\) across 647 samples. Uniqueness is assessed by checking for duplicate structures among the proposals using pymatgen, and Novelty by checking against Materials Project structures in the same way. All S.U.N. criteria were evaluated after MatterSim relaxation, including relaxation of both the proposed candidates and the Materials Project reference phases used for convex-hull construction. Here, reference phases denote competing phases in the corresponding chemical systems that can serve as decomposition products. For band gap, both the predictors used in SMOACS and the band-gap fine-tuning dataset used in MatterGen are based on DFT band-gap labels generated with Materials Project-compatible workflows~\cite{chen2019graph, zeni2023mattergen}. The discrepancies between our DFT band gaps and both source datasets were approximately \( 0.1~\mathrm{eV}\), supporting the \(\pm 0.2~\mathrm{eV}\) tolerance used for DFT-level comparisons (SI section~\ref{appendix__dft_band_gap_consistency}).

\begin{figure}
\begin{center}
\includegraphics[width=0.6\linewidth]{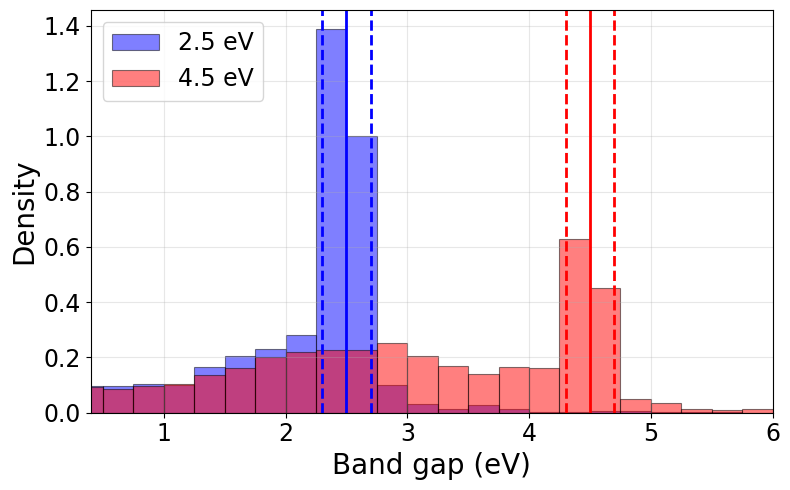}
\end{center}
\caption{Predictor-level band-gap distributions for candidates proposed by SMOACS when targeting $2.5$ (blue) and $4.5$ (red) eV. The purple shading represents the visual overlap of the two semi-transparent histograms. All values are predictions from the property predictor used in SMOACS. Solid lines indicate the target band gaps, and dashed lines the target windows. }
\label{fig__bg_ml_distribution}
\end{figure}

Table~\ref{table__mattergen_bandgap} highlights complementary strengths of the two approaches. MatterGen, pretrained on 608{,}000 structures, has a strong prior for stable crystals and more often generates near-hull candidates, yielding higher S.U.N. success rates. SMOACS directly optimizes the band-gap target (Figure~\ref{fig__bg_ml_distribution}) and gives broadly comparable DFT band-gap hit rates, but its formation-energy term does not reliably yield near-hull structures. After DFT validation, many candidates leave the target window, likely reflecting the approximately 0.2 eV predictor error, the additional approximately 0.1 eV mismatch between the MEGNet dataset and the DFT calculations used here, and reduced predictor reliability for unstable candidates, which are underrepresented in the MEGNet training data (SI section~\ref{appendix__bg_discrepancy}). This highlights a key limitation of SMOACS, namely that its performance is bounded by predictor fidelity.
\begin{table}[t]
\centering
\caption{DFT-assessed band-gap targeting benchmark against MatterGen. For each target, 512 generated candidates were relaxed and evaluated by DFT. ``BG SC rate'' is the success rate for the target DFT band-gap criterion. ``S.U.N. SC rate'' is the success rate for the Stability--Uniqueness--Novelty (S.U.N.) screen used for representative candidates~\cite{zeni2023mattergen}. ``BG\& S.U.N.'' is the fraction of relaxed candidates that satisfy both criteria simultaneously.}
\label{table__mattergen_bandgap}
\begin{tabular}{llccc}
\toprule
\makecell{Target\\$E_g$ (eV)} & Method & \makecell{BG\& \\ S.U.N.} & \makecell{BG\\SC rate} & \makecell{S.U.N.\\SC rate} \\
\midrule
\multirow{2}{*}{\makecell{$2.5$\\$\pm0.2$}}
& MatterGen & \underline{\textbf{0.045}} & 0.062 & 0.584 \\
& SMOACS    & 0.012 & 0.049 & 0.150 \\
\midrule
\multirow{2}{*}{\makecell{$3.0$\\$\pm0.2$}}
& MatterGen & 0.033 & 0.061 & 0.537 \\
& SMOACS    & \underline{\textbf{0.037}} & 0.098 & 0.217 \\
\midrule
\multirow{2}{*}{\makecell{$3.5$\\$\pm0.2$}}
& MatterGen & \underline{\textbf{0.041}} & 0.068 & 0.553 \\
& SMOACS    & 0.021 & 0.061 & 0.176 \\
\midrule
\multirow{2}{*}{\makecell{$4.0$\\$\pm0.2$}}
& MatterGen & 0.033 & 0.074 & 0.508 \\
& SMOACS    & \underline{\textbf{0.034}} & 0.058 & 0.236 \\
\midrule
\multirow{2}{*}{\makecell{$4.5$\\$\pm0.2$}}
& MatterGen & \underline{\textbf{0.039}} & 0.059 & 0.496 \\
& SMOACS    & 0.016 & 0.041 & 0.217 \\
\bottomrule
\end{tabular}
\end{table}
Importantly, the joint ``BG\& S.U.N.'' success rate in Table~\ref{table__mattergen_bandgap}, defined as the fraction of relaxed candidates that both fall within the target DFT band-gap window and pass the S.U.N. screen, remains competitive even though SMOACS relies on predictors trained on only 60{,}000 structures. This indicates that SMOACS can propose candidate materials at a joint success rate comparable to that of generative models trained on much larger datasets, despite relying on predictors using a moderately sized training dataset.   

\begin{figure}[htpb]
\begin{center}
\includegraphics[width=0.92\linewidth]{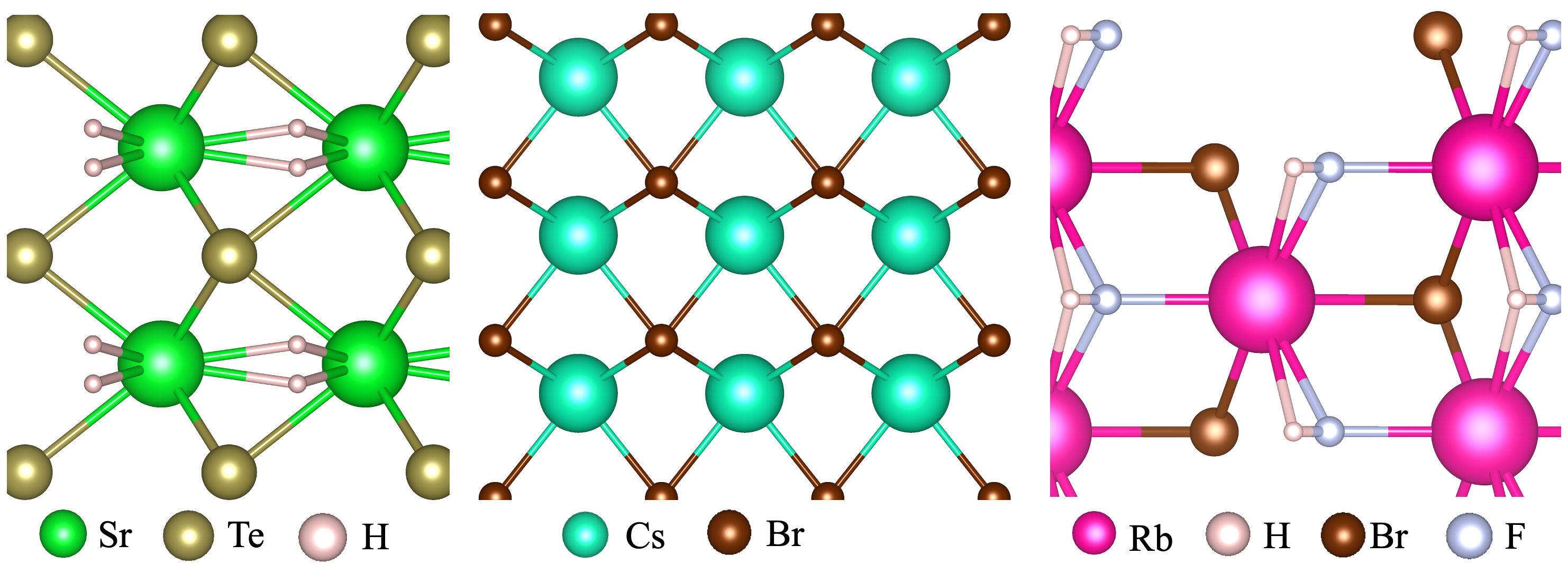}
\end{center}
\caption{Representative candidates that satisfy both the target DFT band-gap window and the S.U.N. criteria. From left to right: $\mathrm{SrTeH_2}$ targeted at 2.5 eV ($E^{\mathrm{DFT}}_{\mathrm{hull}} = 0.01$ eV/atom, $E^{\mathrm{DFT}}_{g} = 2.44$ eV), $\mathrm{CsBr}$ targeted at 4.0 eV ($E^{\mathrm{DFT}}_{\mathrm{hull}} = 0.04$ eV/atom, $E^{\mathrm{DFT}}_{g} = 4.00$ eV), and $\mathrm{RbHBrF}$ targeted at 4.5 eV ($E^{\mathrm{DFT}}_{\mathrm{hull}} = 0.00$ eV/atom, $E^{\mathrm{DFT}}_{g} = 4.34$ eV). Visualized with VESTA~\cite{Momma:db5098}.}
\label{fig:bg_sun_candidates}
\end{figure}

Figure~\ref{fig:bg_sun_candidates} shows representative candidates that satisfy both the target DFT band-gap window and the S.U.N. criteria. For these candidates, we used not only Materials Project but also Alexandria Materials Database~\cite{alexandria_db} for reference phase sets of S.U.N. criteria. The Stability component of S.U.N. was determined from the DFT-derived convex-hull energy, $E^{\mathrm{DFT}}_{\mathrm{hull}}$, rather than from MatterSim energies, while Uniqueness and Novelty were evaluated as described above. The examples include near-hull candidates obtained from different band-gap targeting runs: $\mathrm{SrTeH_2}$ for the 2.5 eV target ($E^{\mathrm{DFT}}_{\mathrm{hull}} = 0.01$ eV/atom, $E^{\mathrm{DFT}}_{g} = 2.44$ eV), $\mathrm{CsBr}$ for the 4.0 eV target ($E^{\mathrm{DFT}}_{\mathrm{hull}} = 0.04$ eV/atom, $E^{\mathrm{DFT}}_{g} = 4.00$ eV), and $\mathrm{RbHBrF}$ for the 4.5 eV target ($E^{\mathrm{DFT}}_{\mathrm{hull}} = 0.00$ eV/atom, $E^{\mathrm{DFT}}_{g} = 4.34$ eV).

\begin{figure}[t]
\begin{center}
\includegraphics[width=1.0\linewidth]{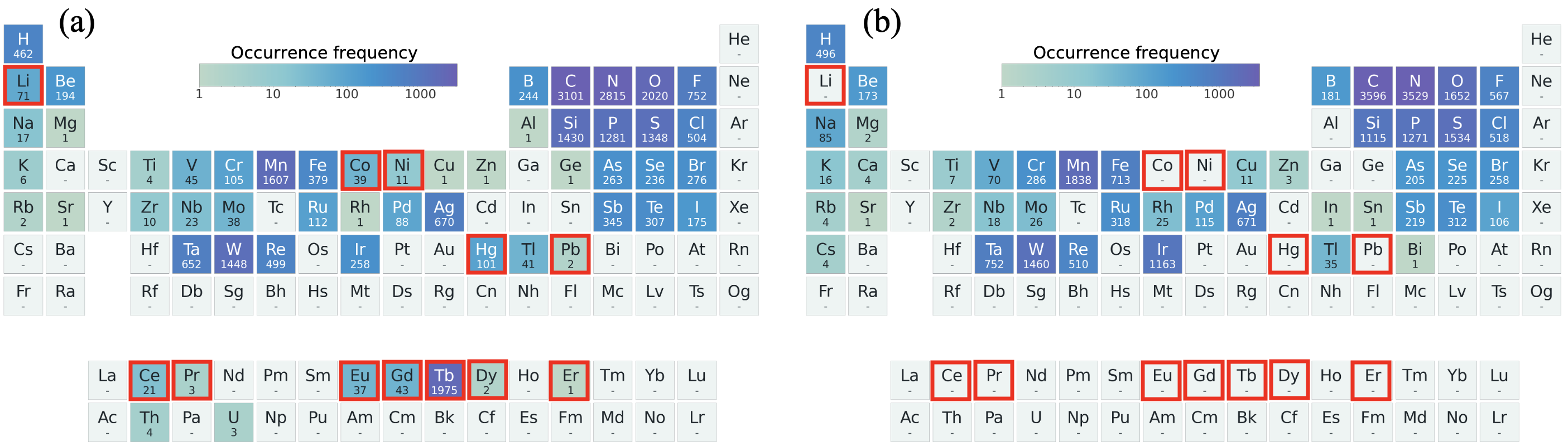}
\end{center}
\caption{Effect of the additional element-exclusion mask in the half-metal design task. Panel (a) shows the element frequencies obtained with only the oxidation-number restriction. Panel (b) shows the result after additionally excluding toxic and supply-risk-relevant elements, including rare-earth elements. Red boxes highlight species that disappear once the added mask is applied.}
\label{fig:mask_effect}
\end{figure}

\subsection{Constrained SMOACS--DFT search for half-metal candidates}
\label{sec__half_metal}

Finally, we use SMOACS here as the front end of a high-throughput DFT screening pipeline to demonstrate that predictor-guided crystal optimization can be tailored to a constraint-rich, application-oriented materials-design task using off-the-shelf property predictors. As a concrete case study for this demonstration, we focus on half-metal-candidate search. Half-metals are attractive spintronic electrode candidates because highly spin-polarized carriers at the Fermi energy (\(E_F\)) are desirable for spin injection and magnetoresistive devices~\cite{jourdan2014direct, elphick2021heusler, katsnelson2008half}. Although Co-based Heusler alloys are widely studied reference systems in the half-metal and spintronic-materials literature~\cite{elphick2021heusler}, Co is identified as supply-risk-relevant in multiple official critical-material assessments~\cite{aist2008rare, ec_jrc_rmis_critical_raw_materials_2023, usgs_final_2025_critical_minerals, doe_2023_critical_materials_list}. Motivated by this context, we used the stricter project-defined exclusion set as a stress test for constrained materials discovery, asking whether SMOACS can identify DFT-level half-metal candidates while excluding a minimal RoHS-relevant heavy-metal subset~\cite{rohs2011} and selected supply-risk-relevant elements, including Co. Because no direct half-metallicity predictor was available within the pretrained models used here, we configured this task as a proxy-guided screening problem to demonstrate the flexibility of SMOACS in combining related property predictors with differentiable loss terms.

To implement this design task, we combined the oxidation-number mask with an additional exclusion mask removing a minimal RoHS-relevant heavy-metal subset~\cite{rohs2011} ($\mathrm{Pb}$, $\mathrm{Cd}$, and $\mathrm{Hg}$), a project-defined non-rare-earth supply-risk-motivated subset ($\mathrm{Co}$, $\mathrm{Li}$, $\mathrm{Ni}$, $\mathrm{Ga}$, and $\mathrm{Pt}$), and rare-earth elements ($\mathrm{Sc}$, $\mathrm{Y}$, and the lanthanides). Additionally, to promote more realistic compositions, we added an \textit{element-reduction loss} term that discourages candidates with too many distinct elements, implemented as an $L_1$ penalty on the site-wise atomic distributions. The proposed structures were subsequently passed through S.U.N. filtering and validated by spin-resolved DFT density-of-states (DOS) calculations using a Materials Project-compatible workflow.

% To implement this practical design task, we combined the oxidation-number mask with an additional exclusion mask removing a minimal RoHS-relevant heavy-metal subset~\cite{rohs2011}($\mathrm{Pb}$, $\mathrm{Cd}$, and $\mathrm{Hg}$), a project-defined supply-risk-motivated subset($\mathrm{Co}$, $\mathrm{Li}$, $\mathrm{Ni}$, $\mathrm{Ga}$, and $\mathrm{Pt}$), and rare-earth elements ($\mathrm{Sc}$, $\mathrm{Y}$, and the lanthanides). The supply-risk-motivated subset was selected referenced withofficial criticality frameworks based on non-rare-earth elements shared across the EU 2023, USGS 2025, and DOE 2023 energy-critical lists. First, to favor energetic stability, we minimized the MatterSim-predicted energy~\cite{yang2024mattersim}. Second, because no direct half-metal predictor was available, we used two soft proxy terms from pretrained ALIGNN band-gap and magnetic-moment models~\cite{choudhary2021atomistic}, favoring metallic or near-metallic electronic character and encouraging an integer total magnetic moment. The latter is a useful empirical proxy motivated by the Slater--Pauling behavior widely observed in half-metallic Heusler alloys~\cite{galanakis2006electronic}. Finally, to promote more realistic compositions, we added an \textit{element-reduction loss} term that discourages candidates with too many distinct elements, implemented as an $L_1$ penalty on the site-wise atomic distributions. The proposed structures were subsequently passed through S.U.N. filtering and spin-resolved DFT validation using a Materials Project-compatible workflow.

Figure~\ref{fig:mask_effect} shows the effect of the additional exclusion mask. With only the oxidation-number restriction, a number of undesirable species still appear in the optimized proposals. Once the extra mask is activated, the highlighted forbidden elements disappear from the proposal set: $\mathrm{Li}$, $\mathrm{Co}$, $\mathrm{Ni}$, $\mathrm{Hg}$, $\mathrm{Pb}$, and several lanthanides are present in Figure~\ref{fig:mask_effect}(a) but absent in Figure~\ref{fig:mask_effect}(b). Importantly, the restriction acts as a hard design rule applied during optimization itself, not as a weak preference or a post hoc filtering step.

\begin{figure}[hbtp]
\begin{center}
\includegraphics[width=0.6\linewidth]{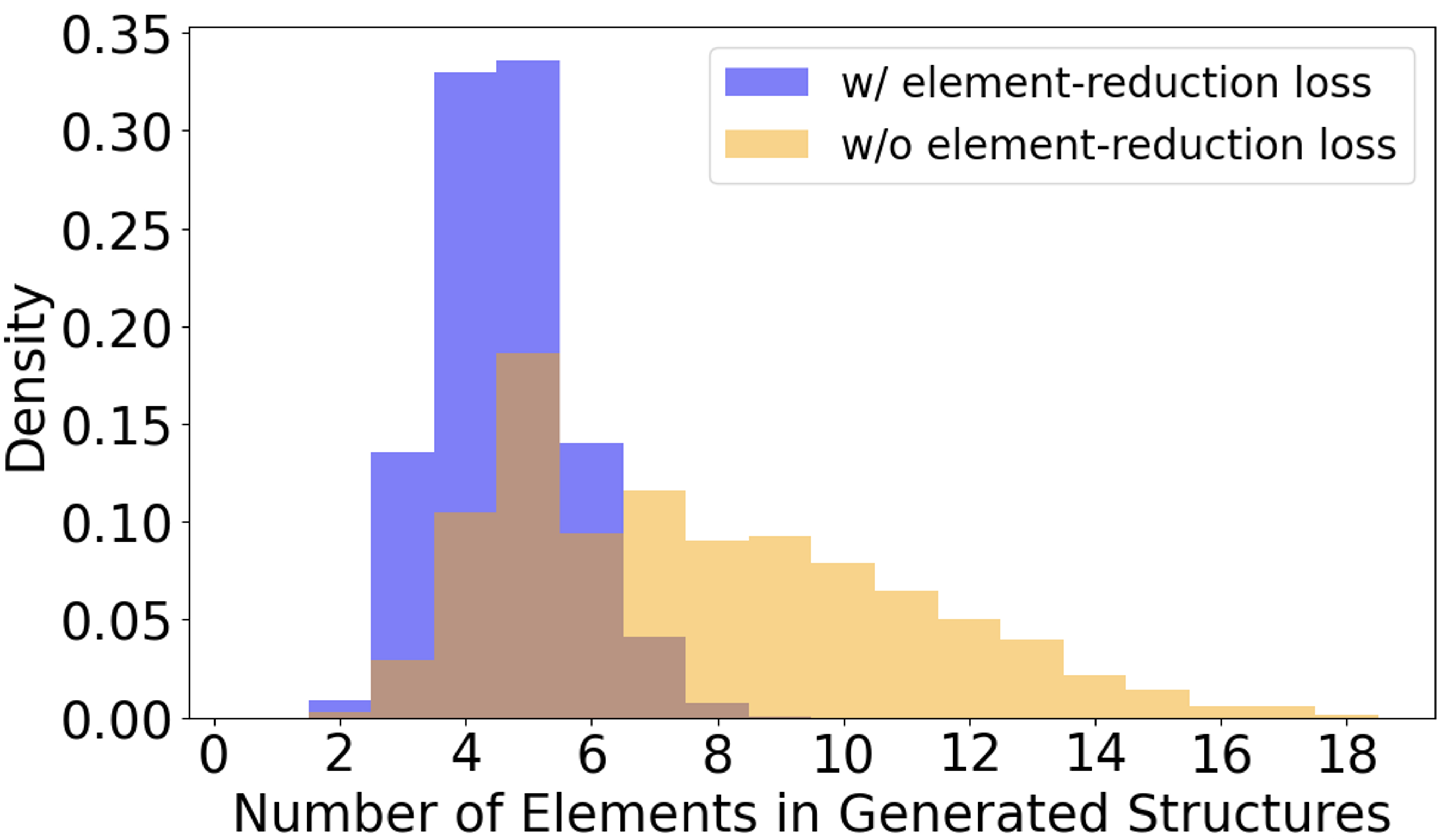}
\end{center}
\caption{Effect of element-reduction loss on compositional complexity in the half-metal design task. Adding element-reduction loss shifts the proposal distribution toward structures containing fewer distinct elements.}
\label{fig:element_loss_effect}
\end{figure}

Figure~\ref{fig:element_loss_effect} then shows the role of element-reduction loss. Without this term, the optimization tends to exploit high compositional complexity. Adding element-reduction loss shifts the distribution toward substantially simpler candidates, concentrating the proposals around roughly four to six elements per structure instead of the much broader range seen without the term. This is a practically useful outcome: reducing the number of constituent elements can simplify synthesis and lower the risk of forming competing phases. More broadly, it demonstrates that SMOACS can incorporate differentiable preferences that are not themselves property predictors, thereby allowing project-specific criteria for candidate selection to be encoded directly into the objective.

\begin{figure}[hbtp]
\begin{center}
\includegraphics[width=0.60\linewidth]{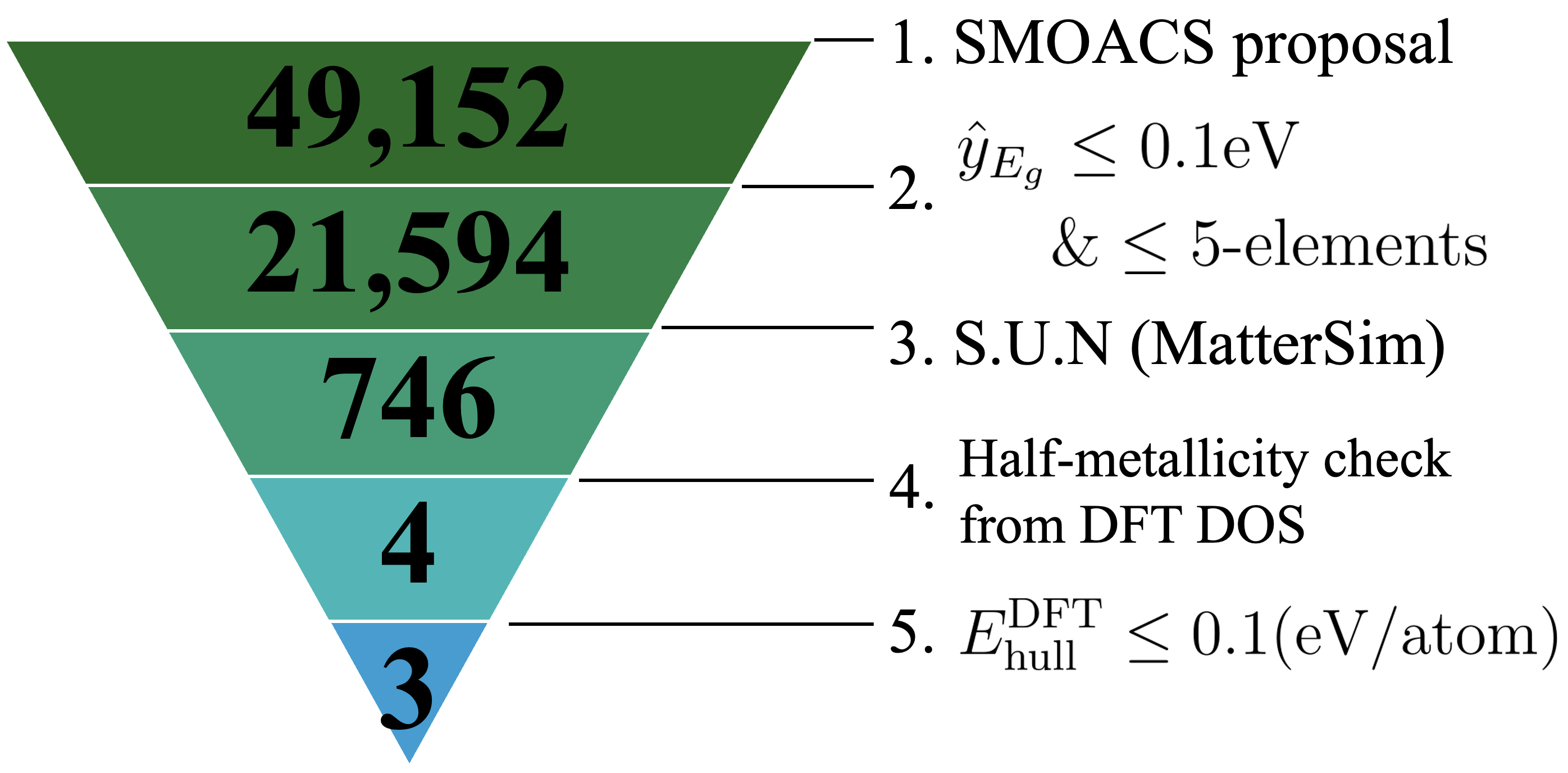}
\end{center}
\caption{Screening workflow for SMOACS-proposed candidates. Starting from 49,152 candidates, materials were sequentially filtered by predicted band gap and number of elements, MatterSim-based S.U.N. criteria, DFT spin-resolved DOS half-metallicity, and DFT convex-hull stability.}
\label{fig:half_metal_workflow}
\end{figure}

\begin{figure}[hbtp]
\begin{center}
\includegraphics[width=0.95\linewidth]{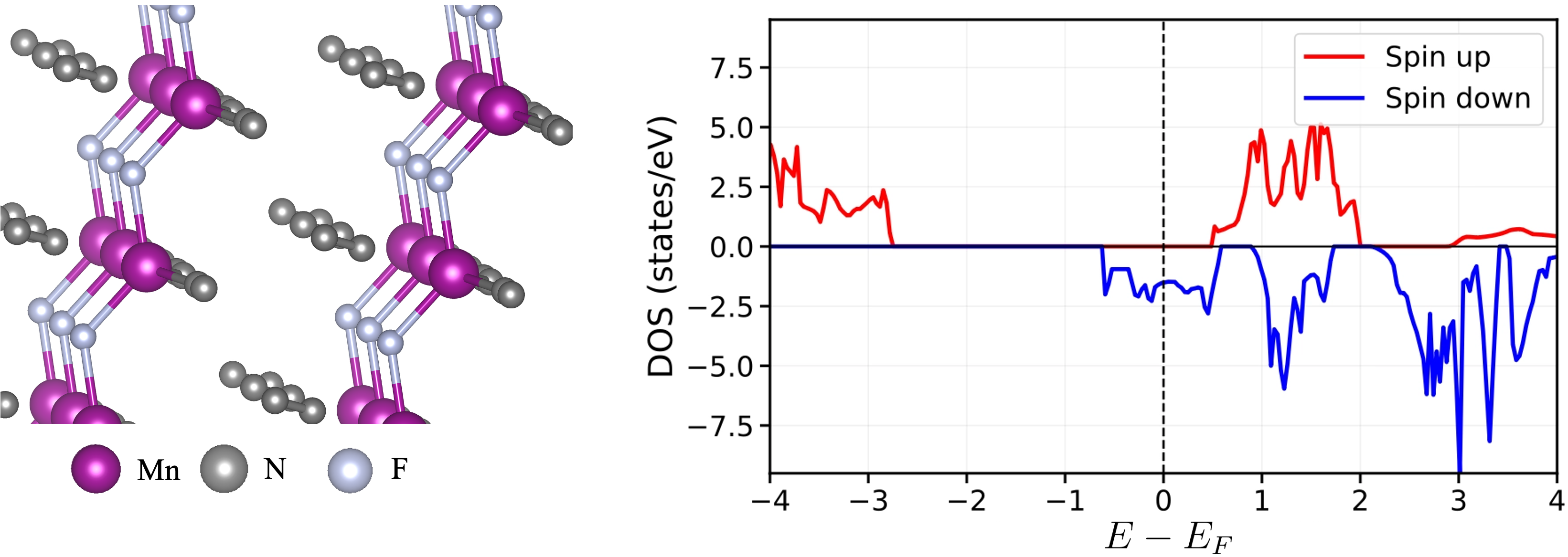}
\end{center}
\caption{Structure and spin-resolved DOS of the representative final half-metal proposal, $\mathrm{MnN_4F}$. The down-spin channel is metallic while the up-spin channel is gapped. The candidate has $E_{\mathrm{hull}}^{\mathrm{DFT}} = 0.06$ eV/atom and a magnetic moment of $4.00~\mu_{\mathrm{B}}$ per formula unit.}
\label{fig:half_metal_candidates}
\end{figure}

The screening procedure shown in Figure~\ref{fig:half_metal_workflow} starts from 49,152 materials proposed by SMOACS. After filtering by the predicted band gap, $\hat{y}_{E_g} \leq 0.1$ eV, and by elemental complexity, requiring five or fewer constituent elements, 21,594 candidates remained. We next applied the MatterSim~\cite{yang2024mattersim}-based S.U.N. screening protocol as in Section~\ref{sec__comparison_with_mattergen}, which reduced the set to 746 candidates. DFT spin-resolved DOS calculations then confirmed half-metallicity for 4 candidates. Finally, to apply a stricter S.U.N. stability assessment, we computed DFT convex-hull energies using reference phases from both the Materials Project and the Alexandria Materials Database, rather than the Materials Project alone. Three candidates were found to satisfy \(E_{\mathrm{hull}}^{\mathrm{DFT}} \leq 0.1\) eV/atom.

Figure~\ref{fig:half_metal_candidates} shows the representative final proposal, $\mathrm{MnN_4F}$. The structure consists of $\mathrm{MnN_2F}$-based layers containing embedded molecular $\mathrm{N_2}$ units. As shown in the spin-resolved DOS, the down-spin channel is metallic whereas the up-spin channel is gapped, indicating half-metallicity. The candidate lies near the convex hull with $E_{\mathrm{hull}}^{\mathrm{DFT}} = 0.06$ eV/atom and has an integer magnetic moment of $4.00~\mu_{\mathrm{B}}$ per formula unit. The other two final candidates are minor configurational variants that mainly differ in the arrangement of the $\mathrm{N_2}$ units.

%
%
% -------------------------------------------------------
% -------------------------------------------------------
% -------------------------------------------------------
%
%

\section{Discussion}\label{sec__discussion}

Practical materials design requires more than proposing crystal structures, as target properties must be optimized together with task-specific constraints. These results indicate that predictor-guided optimization can address this regime by separating predictor training from task specification. In SMOACS, property predictors provide differentiable guidance, while site-wise masks, template-based initialization, and task-specific loss terms define the feasible search space and encode design preferences. Constraints expressible as masks or structural bounds can therefore be enforced during optimization rather than introduced only as post hoc filters. 

The perovskite benchmark shows the value of this formulation. By enforcing charge neutrality and perovskite geometry during optimization, and by combining band-gap and formation-energy objectives with a tolerance-factor term, SMOACS satisfied the full specification more reliably than FTCP, CDVAE, and TPE. The $3\times3\times3$-cell perovskite (135 atoms) experiment showed that the method remains effective even in a roughly 13{,}500-dimensional search space, suggesting applicability to larger doped crystal structures. In the comparison with MatterGen, the generative model benefited from a strong prior learned from a larger training dataset, especially for proposing candidates that remain near the convex hull. Even so, SMOACS achieved a comparable rate of proposals that simultaneously satisfied the DFT band-gap target and the S.U.N. criterion, despite relying on predictors trained on far fewer structures. This result is important because it suggests that, even when suitable off-the-shelf predictors or several-hundred-thousand-structure training sets are unavailable, practical materials design can be enabled by training task-specific predictors on medium-sized datasets.

The half-metal task further illustrates the flexibility of the framework. Additional element-exclusion masks removed prohibited species during optimization, and the element-reduction loss shifted proposals toward simpler compositions. Combined with surrogate objectives for band gap, energy, and magnetic moment, these choices produced a small set of DFT-supported half-metal candidates after stability and spin-resolved density-of-states screening. Note that this experiment is a proof-of-concept demonstration rather than a complete validation of spintronic materials, because magnetic-ordering energetics, electronic-structure validation, and finite-temperature effects were not examined exhaustively.

Several limitations remain. Because SMOACS relies on surrogate predictors, high predictor-level success does not always translate into high DFT hit rates. This limitation may become less severe as property prediction continues to advance rapidly through improved datasets, architectures, and training strategies. Moreover, although \(E_{\mathrm{hull}}\) is useful as a 0~K thermodynamic stability indicator, it does not by itself capture synthesis-relevant factors such as finite-temperature stability, plausible precursors, synthesis routes, or processing conditions~\cite{kovnir2021predictive, sun2025critical, bartel2018physical, szymanski2023autonomous}. Future work should therefore combine SMOACS with uncertainty-aware objectives~\cite{lakshminarayanan2017simple}, ensemble predictors, active-learning loops with DFT feedback~\cite{lookman2019active, yang2024mattersim}, and synthesis-aware screening signals~\cite{szymanski2023autonomous}.

Overall, these results establish SMOACS as a modular front end of a high-throughput DFT pipeline for constraint-rich materials design. Its main value is not to replace generative models or black-box optimization methods, but to provide a way to translate diverse design specifications into optimized candidate structures. Importantly, SMOACS is not limited to settings in which suitable off-the-shelf predictors are already available; task-specific predictors trained on medium-sized datasets can still provide sufficient guidance for efficient materials design. As property predictors and differentiable physics-based objectives improve, the same framework can be extended to broader materials classes and increasingly realistic design criteria.

%
%
% -------------------------------------------------------
% -------------------------------------------------------
% -------------------------------------------------------
%
%

\section{Methods}\label{sec:methods_npj}

\subsection{Problem formulation}

SMOACS directly optimizes a crystal representation using gradients from pretrained property predictors, rather than optimizing a latent variable of a separate generator. Let $f_{*}$ denote a fixed predictor and $\mathcal{L}$ a task-specific objective. The crystal parameters $\boldsymbol{\theta}_{\mathrm{crystal}}$ are updated as
\begin{equation}
\boldsymbol{\theta}_{\mathrm{crystal}}
\leftarrow
\boldsymbol{\theta}_{\mathrm{crystal}}
-
\eta
\nabla_{\boldsymbol{\theta}_{\mathrm{crystal}}}
\mathcal{L}\!\left(f_{*}(\boldsymbol{\theta}_{\mathrm{crystal}})\right).
\label{eq__methods_grad_update}
\end{equation}
We represent each crystal by
\begin{align}
\boldsymbol{\theta}_{\mathrm{crystal}}
&=
\{\boldsymbol{l},\boldsymbol{C},\boldsymbol{A},\boldsymbol{o}\},\\
\boldsymbol{l} \in \mathbb{R}^{6}&, \
\boldsymbol{C} \in \mathbb{R}^{N\times 3}, \
\boldsymbol{A} \in \mathbb{R}^{N\times K}, \
\boldsymbol{o} \in \mathbb{R}^{D},
\end{align}
where $\boldsymbol{l}$ contains the lattice lengths and angles, and $\boldsymbol{C}$ is the matrix of fractional coordinates for $N$ sites. Continuous variables $\boldsymbol{l}$ and $\boldsymbol{C}$ are optimized directly by backpropagation, whereas discrete element choices are not directly optimized. Instead, we relax them into the continuous atomic distributions in $\boldsymbol{A}$~\cite{xiecrystal, zeni2023mattergen} over $K$ candidate elements. At a fully discretized optimum, each row of $\boldsymbol{A}$ becomes one-hot. This parameterization is applicable both to models that take one-hot elemental identities such as Crystalformer~\cite{taniaicrystalformer} and to models that require continuous site embeddings such as ALIGNN~\cite{Kamal2021alignn}. In the latter case, the site representation is obtained as the weighted average of elemental embeddings under the corresponding row of $\boldsymbol{A}$. We also introduce $\boldsymbol{o}$ as a learnable weight over $D$ oxidation-number patterns. This allows multiple oxidation-state patterns to be considered simultaneously. For example, for a perovskite $\mathrm{ABX_3}$, both $\{A\!:\!+2,\,B\!:\!+4,\,X\!:\!-2\}$ and $\{A\!:\!+1,\,B\!:\!+2,\,X\!:\!-1\}$ can be treated within a unified optimization framework.

\subsection{Constraint handling and optimization procedure}

\paragraph{Site-wise element mask}
SMOACS employs site-wise element masks to enforce charge neutrality and additional project-specific element constraints.
To ensure charge neutrality, we construct site-wise oxidation-number masks from the charge-neutral oxidation-number patterns associated with the template structures. For site $n$ assigned oxidation number $s_n$, let $\boldsymbol{m}(s_n)\in\{0,1\}^{K}$ denote the corresponding binary mask over the $K$ candidate elements. The masked atomic distribution is
\begin{equation}
  \tilde{\boldsymbol{a}}_n(s) = \sigma_T\!\left( \boldsymbol{m}(s_n)\odot \boldsymbol{a}_n \right), \ \
  \sigma_T(z_i)= \frac{\exp(z_i/T)}{\sum_j \exp(z_j/T)}.
  \label{eq__methods_mask_single}
\end{equation}
Here, $\sigma_{T}$ is a temperature-scaled softmax function. The mask is not limited to encoding oxidation-number compatibility. Project-specific constraints can also be incorporated in the same way by setting the mask entries of disallowed elements to zero.

For oxidation-pattern index $d$, let $\boldsymbol{S}_d=(s_{1,d},\dots, s_{N, d})$ and let $\boldsymbol{M}(\boldsymbol{S}_d)\in\{0,1\}^{N\times K}$ be the corresponding stacked mask matrix. The full masked atomic distribution is then
\begin{equation}
\tilde{\boldsymbol{A}}_n(\boldsymbol{A},\boldsymbol{o},T)
=
\sigma_{T,\mathrm{elm}}
\left(
\sum_{d=1}^{D}
\sigma_{T,\mathrm{elm}}\!\left(\boldsymbol{M}(\boldsymbol{S}_d)\odot \boldsymbol{A}\right)\tilde{o}_d
\right),
\label{eq__methods_mask_full}
\end{equation}
where $\tilde{o}_d$ is the $d$-th element of $\tilde{\boldsymbol{o}}=\sigma_T(\boldsymbol{o})$ and $\sigma_{T,\mathrm{elm}}$ is a temperature-scaled softmax $\sigma_{T}$ that normalizes along the element axis, producing sharper distributions at lower temperatures $T$. Oxidation states were taken from the SMACT~\cite{Davies2019}--pymatgen~\cite{ONG2013314} intersection (SI section~\ref{appendix__electrical_neutrality}). To prevent masked-out elements from reappearing numerically, values for allowed elements were clipped below at $10^{-6}$ (SI section~\ref{appendix__oxidation_clip}).

\paragraph{Template-based initialization and structure preservation}

SMOACS uses template-based initialization to avoid poor local minima and preserve a target crystal family. Each template provides an initial lattice, fractional coordinates, and one or more charge-neutral oxidation-number patterns (SI section~\ref{appendix__generating_oxidation_patterns}). Each lattice length is parameterized as a learnable multiplier of the template length and rescaled so that the maximum length equals 1 at initialization. Lattice angles are normalized to the range [0,1] over $30^\circ$--$150^\circ$, and $\boldsymbol{A}$ and $\boldsymbol{o}$ are initialized uniformly. 

In the perovskite benchmark, synthetic $\mathrm{ABX}_3$ unit-cell templates used canonical fractional coordinates $(0.5,0.5,0.5)$ for $\mathrm{A}$-site, $(0,0,0)$ for $\mathrm{B}$-site, and $(0.5,0,0)$, $(0,0.5,0)$, and $(0,0,0.5)$ for the three $\mathrm{X}$ sites, with oxidation patterns fixed to $[+2,+4,-2]$ and $[+1,+2,-1]$. To preserve perovskite geometry, optimization started from templates with perturbed lattice lengths, fixed lattice angles at $90^\circ$, and restricted each fractional coordinate to $\pm \epsilon$ of its canonical position. Optimization can also be restricted to a specified space group (SI section~\ref{appendix___spg_conditioned_generation}).

\paragraph{Optimization schedule, temperature annealing, and mutation.}

The general objective is written as
\begin{equation}
\mathcal{L}_{\mathrm{task}}=\sum_i \lambda_i \mathcal{L}_i,
\label{eq__methods_general_loss}
\end{equation}
where each term can be supplied either by a pretrained predictor or by an explicit differentiable function of the crystal representation.

SMOACS used Adam~\cite{kingma2014adam} with cosine annealing. The softmax temperature decayed linearly from $0.01$ to $0.0001$. Unless otherwise noted, lattice lengths and angles were clipped to 2--10~\AA\ and $30^\circ$--$150^\circ$. To improve exploration, we combined gradient updates with mutation and simulated annealing (SA). During mutation, we retained successful candidates and additional top-ranked candidates. We then replaced some lower-ranked candidates with perturbed copies. Gaussian noise was added to their lattice variables and coordinates, their atomic distributions were reinitialized uniformly, and their oxidation masks were inherited. Separately, the SA step proposed a Gaussian perturbation to either the atomic coordinates or the lattice parameters and accepted or rejected it using a criterion with a temperature schedule, thereby allowing occasional acceptance of worse proposals to escape local minima~\cite{kirkpatrick1983optimization}.

In ALIGNN-based optimization, the graph structure was additionally refreshed multiple times during optimization and again after mutation to reflect structural changes. Detailed parameter settings are provided in SI section~\ref{appendix__hyperparameters}.

\subsection{Task-specific experimental protocols}

\paragraph{Perovskite benchmark.}

The perovskite benchmark in Section~\ref{sec__perov_experiments} used publicly available ALIGNN~\cite{choudhary2021atomistic} and Crystalformer~\cite{taniaicrystalformer} predictors for band gap and formation energy trained on the MEGNet dataset~\cite{chen2019graph}. We also incorporated the Goldschmidt tolerance factor $t$, a commonly used stability factor for perovskite structures, into the objective as an explicit differentiable term~\cite{west2022solid}:
\begin{equation}
t=\frac{r_A+r_X}{\sqrt{2}(r_B+r_X)},
\label{eq__methods_tolerance_factor}
\end{equation}
where $r_A$, $r_B$, and $r_X$ are ionic radii. For mixed-anion perovskites, $r_X$ was taken as the average radius over the three $X$ sites.

The perovskite benchmark optimized three properties---band gap, formation energy, and tolerance factor---under two hard constraints: charge neutrality and preservation of the perovskite geometry. The task loss was
\begin{align}
&\mathcal{L}_{\mathrm{perov}} \nonumber \\ &= L_g + L_f + L_t \nonumber \\
&=
\max\!\left(
\left|y_{E_g}^{t}-\hat{y}_{E_g}\right|-0.04,0
\right)
+\hat{y}_{E_f}
+|t-0.9|,
\label{eq__perov_objective}
\end{align}
where $y_{E_g}^{t}$ is the target band gap, $\hat{y}_{E_g}$ is the predicted band gap, and $\hat{y}_{E_f}$ is the predicted formation energy. SMOACS was configured to perform a total of 200 update operations. Initial SMOACS candidates were random perovskite cells with $\alpha=\beta=\gamma=90^\circ$ and $a, b, c$ sampled uniformly from 2--10~\AA. In 3$\times$3$\times$3 perovskite scaling tests, the lattice-length range was expanded proportionally to 6--30~\AA. 

All perovskite SMOACS runs used 200 update steps, with learning rates for coordinates, lattice variables, atomic distributions and oxidation-pattern weights set to 0.005, 0.002, 0.007 and 0.007 for unit-cell Crystalformer; 0.002, 9.0, $5.0\times10^{-5}$ and $5.0\times10^{-5}$ for unit-cell ALIGNN; $6.7\times10^{-4}$, 0.6, 0.002 and 0.002 for $3\times3\times3$ Crystalformer; and $6.7\times10^{-4}$, 0.01, $4.0\times10^{-5}$ and $4.0\times10^{-5}$ for $3\times3\times3$ ALIGNN. Mutation retained successful candidates and the top 10, 87, 42 and 30\% candidates at steps 100/150, 50/100/150, 120/180 and 50/100/150, respectively.
 
\paragraph{Comparison with MatterGen for band-gap optimization.}

In Section~\ref{sec__comparison_with_mattergen}, SMOACS used ALIGNN predictors as in Section~\ref{sec__perov_experiments}. For template-based initialization, we used crystal structures retrieved from the Materials Project as initial templates. We optimized structures with the element-reduction loss $L_{\mathrm{elm}}$, in addition to $L_g$ and $L_f$. The weights of all loss terms were set to unity:
\begin{align}
\mathcal{L}_{\mathrm{S.U.N.-BG}} = L_g + L_f + L_{\mathrm{elm}}.
\label{eq__methods_matgen_bg_compare}
\end{align}

For the S.U.N. screening in Section~\ref{sec__comparison_with_mattergen}, we computed the energy above the convex hull using MatterSim rather than DFT to reduce the computational cost. Reference phases were taken from the Materials Project and processed using the energy-correction scheme implemented in \texttt{MaterialsProject2020Compatibility} in \texttt{pymatgen}. Duplicate removal and novelty checks were performed using \texttt{StructureMatcher}. Results obtained without \(L_{\mathrm{elm}}\) are provided in SI section~\ref{appendix__mattergen_bandgap_without_l_elm}.

For a fair comparison with MatterGen, which uses 1000 diffusion steps, SMOACS was also run for 1000 update operations: 200 gradient-only iterations followed by 400 iterations with one gradient update and one simulated-annealing step per iteration. This schedule corresponds to 1000 update operations in total. The learning rates for coordinates, lattice variables, atomic distributions and oxidation-pattern weights were $9.7\times10^{-3}$, $2.3\times10^{-6}$, $2.6\times10^{-3}$ and $2.6\times10^{-3}$, respectively. Mutation retained successful candidates and the top 20\% candidates at iterations 150, 300 and 450.

\paragraph{Half-metal proposal protocol.}
The half-metal proposal in Section~\ref{sec__half_metal} was formulated as a general crystal-search task without fixing a crystal family. Template structures containing at most 30 atomic sites were retrieved from the Materials Project and used as the initial structures; in total, 8,192 structures were collected.

We combined the oxidation-number mask with an additional exclusion mask that removed a minimal RoHS-relevant heavy-metal subset~\cite{rohs2011} ($\mathrm{Pb}$, $\mathrm{Cd}$, and $\mathrm{Hg}$), a project-defined supply-risk-motivated subset ($\mathrm{Co}$, $\mathrm{Li}$, $\mathrm{Ni}$, $\mathrm{Ga}$, and $\mathrm{Pt}$), and rare-earth elements ($\mathrm{Sc}$, $\mathrm{Y}$, and the lanthanides). The supply-chain-risk subset was selected as the element-level intersection of (i) the \textit{Rare Metals} report from the National Institute of Advanced Industrial Science and Technology (AIST)~\cite{aist2008rare}, (ii) the EU 2023 critical/strategic raw-material framework~\cite{ec_jrc_rmis_critical_raw_materials_2023}, (iii) the USGS 2025 Critical Minerals List~\cite{usgs_final_2025_critical_minerals}, and (iv) the critical-materials-for-energy subset of the 2023 DOE Critical Materials List~\cite{doe_2023_critical_materials_list}.

For the half-metal-candidate search, we optimized an objective combining energy, band-gap, magnetic-moment, and element-reduction terms. The MatterSim-predicted energy per atom, $\hat{y}_{E_{\mathrm{MS}}}$, biases the search toward low-energy configurations. Because no direct half-metallicity predictor was available with ALIGNN and Crystalformer, pretrained ALIGNN band-gap and magnetic-moment models were used as soft proxies~\cite{choudhary2021atomistic}. The band-gap term favors metallic character by driving the predicted gap toward zero, while the magnetic-moment term encourages an integer total moment, motivated by the Slater--Pauling-type behavior observed in half-metallic Heusler alloys~\cite{galanakis2006electronic}. We also included a non-negativity penalty to keep the predicted magnetic moment within the non-negative range represented in the training data, as well as $L_{\mathrm{elm}}$ to discourage overly complex compositions. The loss terms are

\begin{align}
  L_{E_g \to 0} &= \left|\hat{y}_{E_g}\right|,\quad
  L_{E_{\mathrm{MS}}} = \hat{y}_{E_{\mathrm{MS}}}, \\[2pt]
  \hat{y}_{m,+} &= \max(\hat{y}_{m}, 0),\\
  L_{m,\mathrm{nonneg}} &=
  \max\left(
    \varepsilon_{\mathrm{nonneg}} - \hat{y}_{m}, 0
  \right), \\[2pt]
  L_{m,\mathrm{int}} &=
  \max\left(
    \left|\hat{y}_{m,+} - \operatorname{round}(\hat{y}_{m,+})\right|
    - \varepsilon_{\mathrm{tol}}, 0
  \right).
\end{align}

Here, $\hat{y}_{m}$ denotes the predicted total magnetic moment. The constants $\varepsilon_{\mathrm{nonneg}}$ and $\varepsilon_{\mathrm{tol}}$ were set to 0.01 and 0.05, respectively. The full objective is

\begin{align}
L_{\mathrm{half\text{-}metal}}
&= L_{E_g \to 0}
+ L_{E_{\mathrm{MS}}}
+ L_{m,\mathrm{nonneg}} \nonumber \\[2pt]
&\quad
+ w_{m,\mathrm{int}}L_{m,\mathrm{int}}
+ w_{\mathrm{elm}}L_{\mathrm{elm}}.
\end{align}

Here, $w_{m,\mathrm{int}}$ and $w_{\mathrm{elm}}$ are weighting coefficients. We varied both from 1 to 100, generated 49{,}152 structures, and screened them using the workflow described in Section~\ref{sec__half_metal}.

In the half-metal search, we used 650 optimization iterations. The first 300 iterations used gradient updates only, and the remaining 350 iterations combined each gradient update with one simulated-annealing update, corresponding to 1000 update operations in total. The learning rates for coordinates, lattice variables, atomic distributions and oxidation-pattern weights were $2.4\times10^{-6}$, $1.1\times10^{-6}$, $4.1\times10^{-4}$ and $4.1\times10^{-4}$, respectively. Mutation retained successful candidates and the top 20\% candidates at iterations 162, 324 and 486. Detailed hyperparameters are provided in SI section~\ref{appendix__hyperparameters}.

\subsection{Baseline methods}

In Section~\ref{sec__perov_experiments}, we implemented TPE using Optuna~\cite{10.1145/3292500.3330701} with the same pretrained Crystalformer~\cite{taniaicrystalformer} predictors for band gap and formation energy as in the Crystalformer-based SMOACS setting. TPE optimized the same bounded search variables as SMOACS for 200 trials, with band gap, formation energy, charge neutrality, and the tolerance-factor window used as objective terms; charge neutrality was omitted from the objective in the $3\times3\times3$ setting for computational reasons. 

For the generative baselines in Section~\ref{sec__perov_experiments}, perovskite structures in the MEGNet dataset were augmented by translating their fractional coordinates to increase their fraction from approximately $0.2\%$ to $30\%$, thereby encouraging perovskite generation. FTCP~\cite{REN2022314} used the official implementation and was trained from scratch on this perovskite-augmented MEGNet dataset. The training learning rates were set to $1.0\times10^{-4}$ for the unit-cell setting, and $5.0\times10^{-5}$ for the $3\times3\times3$ setting, respectively. For CDVAE~\cite{xiecrystal}, we used the official implementation, jointly trained it with property predictors on the same perovskite-augmented MEGNet dataset, and optimized the latent variables. The learning rates were $1.0\times10^{-3}$ and $1.0\times10^{-4}$ for the unit-cell and $3\times3\times3$ settings, respectively; latent-space optimization used a learning rate of $1.0\times10^{-3}$ for 800 and 5000 steps, respectively. To keep the FTCP and CDVAE training setups aligned with the atom-count limits used in their respective original papers and official implementations, training was restricted to crystals containing at most 20 atoms. For both generative baselines, hyperparameters were tuned; see SI section~\ref{appendix__Detailed_Experimental_Methodology} for details.

In Section~\ref{sec__comparison_with_mattergen}, we used the public MatterGen implementation with the publicly released pretrained weights. Following the official guide, we generated structures using 1{,}000 diffusion steps with $\gamma=2.0$.

\subsection{DFT calculations}

DFT calculations were performed using the Vienna \textit{Ab initio} Simulation Package (VASP)~\cite{hafner2008ab} version 5.4.4 with the Perdew--Burke--Ernzerhof (PBE) generalized-gradient-approximation (GGA) exchange-correlation functional and projector-augmented-wave method~\cite{PhysRevB.59.1758,PhysRevLett.77.3865,PhysRevB.50.17953}. VASP input files were generated using the \texttt{MPGGADoubleRelaxStaticMaker} workflow in Atomate2~\cite{ganose2025atomate2} version 0.0.12 and pymatgen~\cite{ONG2013314} version 2024.8.9. This workflow implements the Materials Project-compatible PBE/PBE+\textit{U} protocol, in which two sequential structural relaxations are followed by a final static calculation.

\subsection{Software and hardware}
SMOACS was implemented in PyTorch~\cite{NEURIPS2019_bdbca288} and run on an NVIDIA A100 GPU. The official implementations of ALIGNN, Crystalformer, and MatterSim were used with minor modifications to enable end-to-end backpropagation.
%
%
% -------------------------------------------------------
% -------------------------------------------------------
% -------------------------------------------------------
%
%

%
%
% -------------------------------------------------------
% -------------------------------------------------------
% -------------------------------------------------------
%
%

\section*{Data availability statement}

No custom training dataset was generated or used in this study. SMOACS used publicly released pretrained models, whereas the FTCP and CDVAE baselines were trained using the MEGNet dataset~\cite{chen2019graph} as described in the Methods section. Reference phase sets for S.U.N. were obtained from the Materials Project~\cite{10.1063/1.4812323} and the Alexandria Materials Database~\cite{alexandria_db}, with database access performed between January 30 and April 28, 2026. 

\section*{Code availability}

The SMOACS implementation, modified pretrained-predictor interfaces, and scripts used to generate and analyse the reported results are available at \url{https://github.com/AkiraTOSEI} and archived on Zenodo at doi:10.5281/zenodo.20149017. The atomate2- and pymatgen-based workflows used for the DFT calculations are available at \url{https://github.com/AkiraTOSEI/atomate2-dft} and archived on Zenodo at doi:10.5281/zenodo.20133844. The comparison-method code is archived on Zenodo at doi:10.5281/zenodo.20133650. Public pretrained models and third-party packages are described in the Methods section and cited there.

\section*{Supplementary information}

This Supplementary Information file contains the following sections: (A) Supplementary discussion, (B) additional implementation and computational settings, and (C) additional experimental results.

\section*{Author contributions}

A.F. conceived the study, developed the SMOACS framework, implemented the computational workflow, performed the experiments and analyses, prepared the figures and tables, and wrote the manuscript. Y.U., A.K.A.L. and K.S. advised on machine-learning methodology, computational materials analysis and interpretation of the DFT-validated candidates. S.W. supervised the project and contributed to study design, interpretation of the results, and manuscript revision. All authors reviewed and approved the manuscript.

\section*{Acknowledgements}

This work was supported by AIST KAKUSEI project (2024). Some computations were performed in part using the Supercomputer System Miyabi at the Joint Center for Advanced High Performance Computing (JCAHPC), jointly operated by the Information Technology Center, The University of Tokyo, and the Center for Computational Sciences, University of Tsukuba.

\section*{Competing interests}
The authors declare no financial or non-financial competing interests.

%
%
% -------------------------------------------------------
% -------------------------------------------------------
% -------------------------------------------------------
%
%

%
%
% -------------------------------------------------------
% -------------------------------------------------------
% -------------------------------------------------------
%
%

\bibliographystyle{sn-nature}
\bibliography{sn-bibliography}% common bib file

%
%
% -------------------------------------------------------
% -------------------------------------------------------
% -------------------------------------------------------
%
%

%
%
% -------------------------------------------------------
% -------------------------------------------------------
% -------------------------------------------------------
%
%

% Appendix を1段組にする.
\clearpage

\begin{appendices}

\numberwithin{equation}{section} % APPENDIXで式番号を1から振り直すのみ使う
\numberwithin{figure}{section} % APPENDIXで図番号を1から振り直すのみ使う
\numberwithin{table}{section} % APPENDIXで表番号を1から振り直すのみ使う

\section{Supplementary Discussion}

% \subsection{Applicability to property prediction models}  \label{sec__general_pplicability}
% This strategy of using atomic distributions discussed in Section~\ref{sec__SMOACS} is widely applicable to various property prediction models. It readily supports formats such as Crystalformer, where one-hot vectors representing elements are fed into the model. Next, we consider a scenario of using models such as ALIGNN that require atomic representations as input. In this scenario, we treat the inner product of the atomic distribution $\boldsymbol{a}_n$ and the $u$-dimensional representation vector for atoms $\boldsymbol{r}_\mathrm{atom} \ (r_\mathrm{atom} \in \mathbb{R}^{K\times u})$ as the atomic representation. In either case, since the output is connected to the learnable atomic distributions through the chain rule of differentiation, we are able to optimize the atomic distribution through backpropagation.

\subsection{Dataset-level consistency of DFT band gaps}
\label{appendix__dft_band_gap_consistency}

To assess the consistency of our DFT band-gap calculations with the reference data used in this study, we compared the band gaps recomputed with our DFT workflow against two source datasets: the MEGNet dataset used to train the SMOACS predictors and the band-gap source data used for MatterGen training. 

For each dataset, we selected entries with reference band gaps between 0 and 5 eV. This range was divided into ten bins with a width of 0.5 eV, and 50 structures were sampled from each dataset using the same stratified protocol. We then recomputed their band gaps using our DFT workflow and evaluated the absolute difference, 
\(|E_g^{\mathrm{DFT}} - E_g^{\mathrm{data}}|\). The 80th percentile of this difference was 0.138 eV for the MEGNet dataset and 0.116 eV for the MatterGen source data. Both values are below the \(\pm 0.2\) eV window used in the band-gap targeting experiments, supporting this tolerance as a reasonable operational criterion for DFT-level comparisons in the present study. 

\subsection{Band-gap discrepancy between predictor-level optimization and DFT validation} \label{appendix__bg_discrepancy}

The ALIGNN band-gap predictor used in SMOACS has a prediction error of approximately 0.2~eV~\cite{Kamal2021alignn}. In addition, the band gaps in its training dataset and those obtained with the DFT protocol used in this work differ by approximately 0.1~eV on average (see Section~\ref{appendix__dft_band_gap_consistency}). It may also be partly attributable to the instability of many proposed structures: predictors trained on the MEGNet dataset, which is dominated by stable structures, may extrapolate poorly to such unstable candidates. These set a practical limit on the transfer from predictor-level optimization to first-principles validation. Thus, even when SMOACS strongly concentrates optimized candidates around the target band gaps at the predictor level, as shown in Figure~\ref{fig__bg_ml_distribution}, the final DFT-validated values can deviate from the predictor-level optima.

\begin{figure}
\begin{center}
\includegraphics[width=0.8\linewidth]{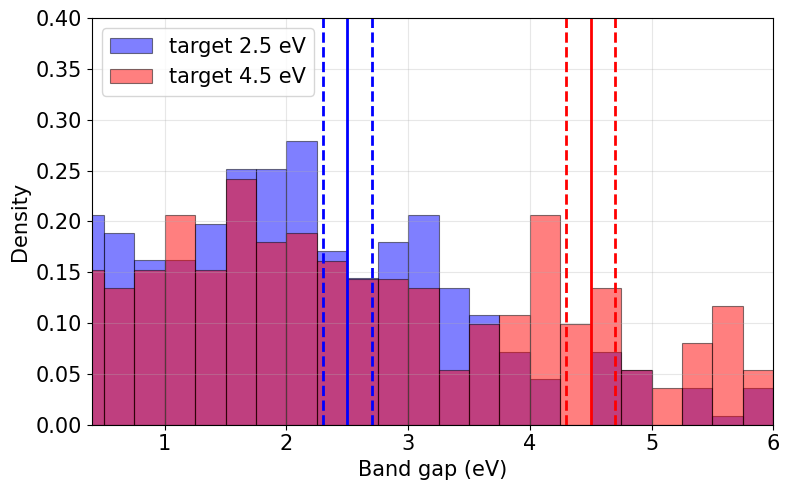}
%\framebox[4.0in]{$\;$}
%\fbox{\rule[-.5cm]{0cm}{4cm} \rule[-.5cm]{4cm}{0cm}}
\end{center}
\caption{Distribution of DFT-validated band gaps for optimized candidates. The broader peaks relative to Figure~\ref{fig__bg_ml_distribution} reflect predictor error and dataset--DFT discrepancies.}
\label{fgr__bg_discrepancy}
\end{figure}

Figure~\ref{fgr__bg_discrepancy} shows the distribution of DFT-validated band gaps for the optimized candidates. The peaks associated with different targets remain separated, indicating that the predictor-guided search preserves target-dependent trends after DFT validation. However, compared with the predictor-level distributions in Figure~\ref{fig__bg_ml_distribution}, the DFT-validated peaks are substantially broader and less sharply concentrated. This broadening reflects the combined effect of predictor error and the dataset--DFT discrepancy, and shows that the final band-gap accuracy of SMOACS is bounded by the fidelity of the surrogate predictor.

\clearpage

\section{Additional implementation and computational settings} \label{appendix__Detailed_Experimental_Methodology}

\subsection{Additional implementation settings for baseline methods}\label{appendix__implement_details}

\subsubsection{Details for TPE} 
Optuna~\cite{10.1145/3292500.3330701} was used for TPE. The number of optimization steps, the range of the crystal axis lengths $a, b, c$, and the angles $\alpha, \beta, \gamma$ were set to the same values as those used for SMOACS. TPE required separate settings for each objective: band gap, formation energy, and electrical neutrality. The objectives for band gap and formation energy were adopted from Equation~\ref{eq__perov_objective}. Additionally, we implemented a binary objective function that assigns a value of $0$ if electrical neutrality is achieved and $1$ otherwise. For $t$, TPE used another objective function:

\begin{gather}
\mathrm{L}_\text{neutral} = 
\begin{cases}
0 & \text{electrical neutrality} \\
1 & \text{otherwise} \ \ \ \
\end{cases}
, \ \ \
    L^\mathrm{TPE}_{t} = 
    \begin{cases}
    0 &(0.8 \leq t \leq 1.0) \\
    1 & \text{otherwise}
    \end{cases}
\end{gather}

%\begin{gather}
%\mathrm{L}^\mathrm{TPE}_{t} = 
%\begin{cases}
%0 &(0.8 \leq t \leq 1.0) \\
%1 & \text{otherwise}
%\end{cases}
%\end{gather}

\counterwithin{equation}{section} % <-- Now recommended

TPE used $L_g$, $L_{E_f}$, $L_\text{neutral}$ and $L^\mathrm{TPE}_{t}$ as objective functions, respectively.

\subsubsection{Details for FTCP} 
We used the publicly available official implementation of FTCP and trained it on the MEGNet dataset. Following the original FTCP paper and its official implementation, we trained FTCP from scratch using the MEGNet dataset. We tuned the hyperparameters, including the $\texttt{max\_elms}$ parameter (the maximum number of element types in the crystal), the $\texttt{max\_sites}$ parameter (the maximum number of atomic sites in the crystal), and the learning rate. The learning rate was selected from the range $1 \times 10^{-5}$ to $1 \times 10^{-3}$, and the standard deviation of the noise added to the latent variables during inference ($L_p$) was selected from the range $1 \times 10^{-6}$ to $1.0$ by empirical evaluation.

As a result, for the experiments in Section~\ref{sec__perov_experiments}, $\texttt{max\_elms}$, $\texttt{max\_sites}$, and the learning rate were set to 7, 20, and 0.0001, respectively. Following the training conditions of ALIGNN and Crystalformer, we used 60{,}000 data points sampled from the MEGNet dataset for training. Note that the MEGNet dataset contains samples with a larger number of element types and atomic sites than these settings, so we did not utilize all 60,000 training samples in Section~\ref{sec__perov_experiments}. Empirically, using the full dataset did not lead to improved performance in our setting. With this configuration, the mean absolute errors for the band gap and formation energy predictions were 0.41 eV and 0.21 eV/atom, respectively. For the large-scale experiments in Section~\ref{sec__perov_experiments}, we aimed to utilize the entire MEGNet dataset. Therefore, we set the $\texttt{max\_elms}$ and $\texttt{max\_sites}$ to 9 and 300, respectively, and used a learning rate of $5 \times 10^{-5}$.

\subsubsection{Details for CDVAE} 

Following the original CDVAE paper and its official implementation, we jointly trained the generative model and property predictors for each dataset. Following the training conditions of ALIGNN and Crystalformer, we trained our model using 60,000 data points sampled from the MEGNet dataset. We tuned the hyperparameters, including the learning rate for training the generative model and predictors, by selecting from the range $1 \times 10^{-5}$ to $1 \times 10^{-3}$ based on validation loss. As a result, the learning rates selected for the experiments in SI section~\ref{appendix__section__rare_experiment}, unit cell in Section~\ref{sec__perov_experiments}, $3\times3\times3$-size in Section~\ref{sec__perov_experiments} were $1 \times 10^{-3}$, $1 \times 10^{-3}$, and $1 \times 10^{-4}$, respectively.

During latent space optimization, following the original paper, CDVAE optimizes latent variables using gradients, generates crystal structures through denoising, and re-encodes the generated structures into latent space for property prediction, since the property predictors operate in latent space. We found that guiding the optimization solely by minimizing formation energy as in the original paper was insufficient to achieve the band gap targets or satisfy the perovskite structure constraints. Therefore, we incorporated multiple predictors to additionally predict these properties and the perovskite identity. Following the training conditions of ALIGNN and Crystalformer, we use 60,000 data points sampled from the MEGNet dataset for training. To reduce the structural diversity and mitigate training difficulty in the perovskite unit cell experiments, we trained CDVAE using only the crystal structures from the MEGNet dataset that contained 20 atoms or fewer for the experiments in Section~\ref{sec__perov_experiments} and SI section~\ref{appendix__section__rare_experiment}. Empirically, using the full dataset did not lead to improved performance in our setting. We also carefully tuned the learning rate for latent space optimization (gradient-based optimization after training the model) and the number of optimization steps to improve convergence and success rates under fair conditions. The optimization learning rate was selected from the range $1 \times 10^{-5}$ to $1 \times 10^{-2}$, and the number of optimization steps from 200 to 5000, based on the success rate. As a result, the learning rate and step count selected for latent space optimization were $1 \times 10^{-3}$ and 3200 in SI section~\ref{appendix__section__rare_experiment}, $1 \times 10^{-3}$ and 800 for unit cells in Section~\ref{sec__perov_experiments}, and $1 \times 10^{-3}$ and 5000 for $3\times3\times3$-size in Section~\ref{sec__perov_experiments}, respectively.

In Section~\ref{sec__perov_experiments}, we tuned the loss coefficient for the perovskite identity classifier over the set $\{1, 10\}$ and selected 10 as the best value. For $3\times3\times3$-size cells, we additionally introduced a predictor for whether a crystal contains more than 100 atoms, with a loss coefficient also selected from $\{1, 10\}$; 10 was selected as the optimal weight. Ideally, by combining this predictor with the perovskite identity classifier, the model is expected to generate large perovskite structures such as \(3 \times 3 \times 3\) or \(3 \times 3 \times 4\) cells.

To ensure a fair comparison, CDVAE began optimization from the same initial structures as SMOACS in all experiments. In Section~\ref{sec__perov_experiments}, we also confirmed that starting optimization from perovskite structures within the dataset did not significantly affect the results.

It is worth noting that the MEGNet dataset is predominantly electrically neutral. Therefore, we did not explicitly perform optimization for charge neutrality during latent optimization, as the generative model is naturally guided toward electrically neutral structures through learning from this dataset.

\subsection{Additional considerations for oxidation-number masks} \label{appendix__oxidation_clip}

In practice, even after applying the oxidation-number mask, we observe that the atomic distribution values for all permitted elements can decrease during gradient-based optimization and may fall below zero. When this happens, masked-out elements---whose values are fixed at zero---can unintentionally dominate after the softmax, leading to the emergence of disallowed elements and violating charge neutrality.

To prevent this issue, we apply a lower bound clipping to the atomic distribution \emph{before} masking, ensuring that all permitted elements retain a small minimum value (e.g., $10^{-6}$). This guarantees that, even if the permitted elements' values decrease during optimization, they never fall below the values of the masked-out elements (which remain at zero). As a result, the masking mechanism reliably enforces charge neutrality throughout the optimization process.

\subsection{Hyperparameter settings for SMOACS} \label{appendix__hyperparameters}

Tables~\ref{table__hyperparams_core}, Table~\ref{table__hyperparams_extra}, and Table~\ref{table__hyperparams_sa} summarize the hyperparameter settings used in all experiments. Table~\ref{table__hyperparams_core} lists the core optimization parameters, including learning rates for crystal lattice ($\eta_l$), atomic coordinates ($\eta_C$), atomic distribution ($\eta_A$), and weights of oxidation-number patterns ($\eta_O$), as well as the number of graph updates \texttt{num\_g\_upd}. We used the Adam optimizer~\cite{kingma2014adam} with a cosine annealing learning rate schedule. Table~\ref{table__hyperparams_extra} and Table~\ref{table__hyperparams_sa} provide parameters related to mutation and SA strategy.

\paragraph{Mutation strategy.}  Mutation is performed when the optimization step reaches any of the values specified in \texttt{mutation\_steps}. First, all candidates are sorted in ascending order of their loss. They are divided into three groups: (i) candidates that have already met the optimization target, (ii) candidates not in group A but within the top fraction specified by \texttt{top\_N\_rate}, and (iii) the remaining candidates. Groups A and B are kept unchanged. For group C, with probability $1 - \texttt{C\_rate}$, each candidate is replaced by a randomly selected candidate from group A or B. After replacement, Gaussian noise with zero mean and standard deviation \texttt{noise} is added to both the atomic coordinates and the learnable scaled lattice constants. The atomic distribution is re-initialized with a uniform distribution, while the oxidation mask from the replacement candidate is reused. Note that mutation operates within each batch of data, meaning that a larger batch size can be advantageous for effective mutation.

\paragraph{Simulated annealing strategy.} SMOACS first performs gradient-based optimization until \texttt{sa\_start\_step} and then applies one simulated annealing (SA) step after each gradient step. SA proposals perturb only the fractional coordinates, scaled lattice lengths, and scaled lattice angles. At SA step $s$, the temperature is set by a geometric schedule from \texttt{sa\_t0} to \texttt{sa\_tmin},
\[
T_s =
\texttt{sa\_t0}
\left(\frac{\texttt{sa\_tmin}}{\texttt{sa\_t0}}\right)^{s/(S-1)},\] 
with endpoints fixed to \texttt{sa\_t0} and \texttt{sa\_tmin}. For each temperature step, \texttt{sa\_trials} proposals are evaluated. Each proposal samples one move type---coordinates, lattice lengths, or lattice angles---with equal probability. Gaussian noise with standard deviations $\sigma_\text{sa\_coords}$, $\sigma_\text{sa\_lattice}$, and $\sigma_\text{sa\_angle}$ is added to the selected variable. Let $\Delta = \mathcal{L}_{\mathrm{proposal}}-\mathcal{L}_{\mathrm{current}}$, where $\mathcal{L}$ is the scalar \texttt{total\_loss}. The proposal is accepted if $\Delta \le 0$; otherwise it is accepted with probability $\exp(-\Delta/T_s)$. Pure SA returns the best state observed over all SA steps, whereas \texttt{grad\_sa\_hybrid} copies the current state after each one-step SA update back into the trainable parameters. SA records the temperature, current loss, best loss, acceptance rate, and per-trial acceptance flags for each step.

\paragraph{Graph updates in ALIGNN.}  In ALIGNN, bonds are defined using a graph structure. Since the graph is non-differentiable, it cannot be optimized directly. Moreover, as the crystal structure evolves during optimization, nearest-neighbor relationships may change, making it inappropriate to keep using the initial graph structure. To address this, we update the graph structure multiple times (\texttt{num\_g\_upd}) during optimization. Specifically, because the learning rate decays according to a cosine schedule, we synchronize graph updates with a sine schedule---the integral of the cosine function---to align with learning-rate decay. For instance, when conducting 15 graph updates over the course of 200 optimization steps, the updates are performed at steps [8, 16, 25, 33, 41, 49, 58, 67, 77, 86, 97, 108, 121, 136, 155]. In addition, the ALIGNN graph is also updated immediately after each mutation and at the end of the optimization.

\clearpage
\begin{table}[ht]
\centering
\small
\caption{Core hyperparameter settings for each experiment. A dash indicates that the parameter was not used.}
\label{table__hyperparams_core}
\begin{tabular}{l l wc{1.2cm} wc{1.2cm} wc{1.5cm} wc{1.5cm} wc{1.5cm} wc{1.5cm}}
\toprule
Section & Model  & total steps & \texttt{num\_g\_upd}  & $\eta_C$ & $\eta_l$ & $\eta_A$ & $\eta_O$ \\
\midrule
Section~\ref{section__init_experiment}& Crystalformer & 200 & - & 0.005 & 0.002 & 0.007 & 0.007 \\
Section~\ref{sec__perov_experiments}~(unit)& Crystalformer & 200 & - & 0.005 & 0.002 & 0.007 & 0.007 \\
Section~\ref{sec__perov_experiments}~(unit)& ALIGNN & 200 & 41 & 0.002 & 9.0 & $5.0 \times 10^{-5}$ & $5.0 \times 10^{-5}$ \\
Section~\ref{sec__perov_experiments}~($3\times3\times3$)& Crystalformer & 200 & - & $6.7 \times 10^{-4}$ & 0.6 & 0.002 & 0.002 \\
Section~\ref{sec__perov_experiments}~($3\times3\times3$)& ALIGNN & 200 & 41 & $6.7 \times 10^{-4}$ & 0.01 & $4.0 \times 10^{-5}$ & $4.0 \times 10^{-5}$ \\
Section~\ref{sec__comparison_with_mattergen}& ALIGNN & 600 & 46 & $9.7 \times 10^{-3}$ & $2.3 \times 10^{-6}$ & $2.6 \times 10^{-3}$ & $2.6 \times 10^{-3}$ \\
Section~\ref{sec__half_metal}& ALIGNN & 650 & 46 & $2.4 \times 10^{-6}$ & $1.1 \times 10^{-6}$ & $4.1 \times 10^{-4}$ & $4.1 \times 10^{-4}$ \\
Section~\ref{appendix__section__rare_experiment}~(Table~\ref{table__rare_property_tc})& ALIGNN & 200 & 12 & 0.002 & $2.0 \times 10^{-4}$ & $2.0 \times 10^{-5}$ & $2.0 \times 10^{-5}$ \\
Section~\ref{appendix__section__rare_experiment}~(Table~\ref{table__rare_property_bg})& Crystalformer & 200 & - & 0.005 & 0.002 & 0.007 & 0.007 \\
Section~\ref{appendix__section__rare_experiment}~(Table~\ref{table__rare_property_bg})& ALIGNN & 200 & 46 & $7.0 \times 10^{-4}$ & $7.0 \times 10^{-4}$ & $3.0 \times 10^{-4}$ & $3.0 \times 10^{-4}$ \\
Section~\ref{section__cross_dataset}~(Table~\ref{table__smoacs_ali_megnet-bg_jarvis_e_full})& ALIGNN & 200 & 46 & $7.0 \times 10^{-4}$ & $7.0 \times 10^{-4}$ & $3.0 \times 10^{-4}$ & $3.0 \times 10^{-4}$ \\
Section~\ref{section__cross_dataset}~(Table~\ref{table__smoacs_ali_megnet-Ef_jarvis_BM})& ALIGNN & 200 & 46 & $3.0 \times 10^{-4}$ & 0.001 & $2.0 \times 10^{-5}$ & $2.0 \times 10^{-5}$ \\

\bottomrule
\end{tabular}
\end{table}
\begin{table}[ht]
\centering
\caption{Additional hyperparameter settings for mutation including batch size (\texttt{b\_size}).}
\label{table__hyperparams_extra}
\begin{tabular}{l l c c c c c}
\toprule
Section & Model & \texttt{b\_size} & \texttt{mutation\_steps} & \texttt{top\_N\_rate} & \texttt{noise} & \texttt{C\_rate} \\
\midrule
Section~\ref{section__init_experiment}& Crystalformer & 256 & [100, 150] & 0.1 & $7.0 \times 10^{-5}$ & 0.05 \\  
Section~\ref{sec__perov_experiments}~(unit)& Crystalformer & 256 & [100, 150] & 0.1 & $7.0 \times 10^{-5}$ & 0.05 \\
Section~\ref{sec__perov_experiments}~(unit)& ALIGNN & 256 & [50, 100, 150] & 0.87 & $1.0 \times 10^{-5}$ & 0.59 \\
Section~\ref{sec__perov_experiments}~($3\times3\times3$)& Crystalformer & 32 & [120, 180] & 0.42 & 0.02 & 0.27 \\
Section~\ref{sec__perov_experiments}~($3\times3\times3$)& ALIGNN & 32 & [50, 100, 150] & 0.3 & $3.0 \times 10^{-6}$ & 0.2 \\
Section~\ref{sec__comparison_with_mattergen}& ALIGNN & 128 & [150, 300, 450] & 0.20 & $5.3 \times 10^{-4}$ & 0.020 \\
Section~\ref{sec__half_metal}& ALIGNN & 150 & [162, 324, 486] & 0.20 & $1.2 \times 10^{-8}$ & 0.020 \\
Section~\ref{appendix__section__rare_experiment}~(Table~\ref{table__rare_property_tc})& ALIGNN & 256 & [50, 100, 150] & 0.31 & $5.0 \times 10^{-5}$ & 0.02 \\
Section~\ref{appendix__section__rare_experiment}~(Table~\ref{table__rare_property_bg})& Crystalformer & 256 & [100, 150] & 0.1 & $7.0 \times 10^{-5}$ & 0.05 \\
Section~\ref{appendix__section__rare_experiment}~(Table~\ref{table__rare_property_bg})& ALIGNN & 256 & [50, 100, 150] & 0.3 & $1.0 \times 10^{-5}$ & 0.2 \\
Section~\ref{section__cross_dataset}~(Table~\ref{table__smoacs_ali_megnet-bg_jarvis_e_full})& ALIGNN & 256 & [120] & 0.48 & $3.0 \times 10^{-6}$ & 0.02 \\
Section~\ref{section__cross_dataset}~(Table~\ref{table__smoacs_ali_megnet-Ef_jarvis_BM})& ALIGNN & 256 & [50, 100] & 0.3 & $3.0 \times 10^{-6}$ & 0.1 \\

\bottomrule
\end{tabular}
\end{table}
\begin{table}[ht]
\centering
\caption{Simulated-annealing hyperparameter settings. A dash indicates that the parameter was not used.}
\label{table__hyperparams_sa}
\begin{tabular}{l l wc{1cm} wc{1cm} wc{1cm} wc{1.1cm} wc{1.1cm} wc{1.1cm} wc{1.1cm}}
\toprule
Section & Model
& \texttt{sa\_start\_step}
& \texttt{sa\_t0}
& \texttt{sa\_tmin}
& $\sigma_{\text{sa\_coords}}$
& $\sigma_{\text{sa\_lattice}}$
& $\sigma_{\text{sa\_angle}}$ \\
\midrule
Section~\ref{section__init_experiment}& Crystalformer & - & - & - & - & - & -  \\
Section~\ref{sec__perov_experiments}~(unit)& Crystalformer & - & - & - & - & - & -  \\
Section~\ref{sec__perov_experiments}~(unit)& ALIGNN & - & - & - & - & - & -  \\
Section~\ref{sec__perov_experiments}~($3\times3\times3$)& Crystalformer & - & - & - & - & - & -  \\
Section~\ref{sec__perov_experiments}~($3\times3\times3$)& ALIGNN & - & - & - & - & - & -  \\
Section~\ref{sec__comparison_with_mattergen}& ALIGNN & 200 & 1.3 & $6.4 \times 10^{-5}$  & $4.0 \times 10^{-3}$ & $4.5 \times 10^{-4}$ & $2.5 \times 10^{-3}$ \\
Section~\ref{sec__half_metal}& ALIGNN & 300 & 0.18 & $1.2 \times 10^{-3}$  & $8.8 \times 10^{-3}$ & $1.0 \times 10^{-3}$ & $4.3 \times 10^{-2}$ \\
Section~\ref{appendix__section__rare_experiment}~(Table~\ref{table__rare_property_tc})& ALIGNN & - & - & - & - & - & -  \\
Section~\ref{appendix__section__rare_experiment}~(Table~\ref{table__rare_property_bg})& Crystalformer & - & - & - & - & - & -  \\
Section~\ref{appendix__section__rare_experiment}~(Table~\ref{table__rare_property_bg})& ALIGNN & - & - & - & - & - & -  \\
Section~\ref{section__cross_dataset}~(Table~\ref{table__smoacs_ali_megnet-bg_jarvis_e_full})& ALIGNN & - & - & - & - & - & -  \\
Section~\ref{section__cross_dataset}~(Table~\ref{table__smoacs_ali_megnet-Ef_jarvis_BM})& ALIGNN & - & - & - & - & - & -  \\
\bottomrule
\end{tabular}
\end{table}

%\clearpage

\subsection{Generation of oxidation number patterns} \label{appendix__generating_oxidation_patterns}

In SMOACS, realistic oxidation number patterns are generated based on the compositions of initial crystal structures. Here, we explain this using $\mathrm{RuN}$ (mp-1009770). According to \verb|icsd_oxidation_state| in pymatgen, ruthenium (Ru) and nitrogen (N) can adopt oxidation numbers of $\{+2, +3, +4, +5, +6\}$ and $\{+1, +3, +5, -1, -2, -3\}$, respectively. Therefore, charge neutrality in $\mathrm{RuN}$ is achieved when the oxidation number combinations for Ru and N are $(+2, -2)$ or $(+3, -3)$. Consequently, when using $\mathrm{RuN}$ (mp-1009770) as the initial structure, oxidation number combination patterns of $(+2, -2)$ and $(+3, -3)$ are obtained, and corresponding masks are generated for each.

In Section~\ref{sec__comparison_with_mattergen} and Section~\ref{sec__half_metal}, we used a larger set of oxidation-number patterns, regardless of the elements at each site. We enumerated combinations of allowed oxidation numbers over the sites and retained those whose total charge was zero. Up to 100 such patterns were generated. 

To consider a broader range of oxidation number combinations, we utilized the intersection of oxidation numbers rom \texttt{SMACT} and \texttt{icsd}, as listed in Table~\ref{table__ox_state1}. It should be noted that even when generating oxidation number patterns from "smact" and "icsd", charge neutrality is maintained by applying site-specific elemental constraints using the oxidation numbers in the "Ours" column of Table~\ref{table__ox_state1}.

\subsection{Charge neutrality} \label{appendix__electrical_neutrality}

In assessing charge neutrality, a compound was considered neutral if the sum of the oxidation numbers for the atoms at each site equaled zero. For example, $\mathrm{Fe_3O_4}$ is electrically neutral because the configuration $[\mathrm{Fe, Fe, Fe, O, O, O, O}]$ can assume oxidation numbers of $[+2,+3,+3,-2,-2,-2,-2]$ that sum to zero. A previous study~\cite{xiecrystal} employed SMACT~\cite{Davies2019} to assess charge neutrality. However, SMACT includes some rare oxidation numbers, such as the +7 state of chlorine, which may be less reliable for our purposes. We restricted our analysis to commonly occurring oxidation numbers, selecting those found at the intersection of SMACT and pymatgen. A list of the elements and their corresponding oxidation numbers employed in this study is shown in Table~\ref{table__ox_state1}, Table~\ref{table__ox_state2}, and Table~\ref{table__ox_state3}. In these tables, the 'SMACT' indicates oxidation numbers from \verb|smact.Element|. The 'icsd' and 'common' indicate oxidation numbers from \verb|icsd_oxidation_state| and \verb|common_oxidation_states| in \verb|pymatgen.core.periodic_table.Element|, respectively. 'Ours' represents the oxidation numbers we used in this paper.

\begin{table}[h]
\caption{List of oxidation numbers from Hydrogen (H) to Calcium (Ca).}
\centering
\begin{tabular}{wl{0.2cm}wl{0.2cm}wc{4.3cm}ccc}
\hline
Z&Elm&SMACT&icsd&common&Ours\\ \hline
1 &H&\{ $-1$, $1$\}&\{$-1$, $1$\}&\{$-1$, $1$\}&\{$-1$, $1$\} \\  \hdashline
2 &He& \{ \}& \{ \}& \{ \}& \{ \} \\  \hdashline
3 &Li& \{$1$\}& \{$1$\}& \{$1$\}& \{$1$\} \\  \hdashline
4 &Be&\{ $1$, $2$\}&\{$2$\}& \{$2$\}& \{$2$\} \\  \hdashline
5 &B&\{ $1$, $2$, $3$\}&\{$-3$, $3$\}&\{$3$\}& \{$3$\} \\  \hdashline
6 &C&\{ $-4$, $-3$, $-2$, $-1$, $1$, $2$, $3$, $4$\}&\{$-4$, $-3$, $-2$, $2$, $3$, $4$\}&\{$-4$, $4$\}&\{$-4$, $4$\} \\  \hdashline
7 &N&\{ $-3$, $-2$, $-1$, $1$, $2$, $3$, $4$, $5$\}&\{$-3$, $-2$, $-1$, $1$, $3$, $5$\}&\{$-3$, $3$, $5$\}&\{$-3$, $3$, $5$\} \\  \hdashline
8 &O&\{ $-2$, $-1$, $1$, $2$\}&\{$-2$\}& \{$-2$\}& \{$-2$\} \\  \hdashline
9 &F& \{$-1$\}& \{$-1$\}& \{$-1$\}& \{$-1$\} \\  \hdashline
10 &Ne& \{ \}& \{ \}& \{ \}& \{ \} \\  \hdashline
11 &Na&\{ $-1$, $1$\}&\{$1$\}& \{$1$\}& \{$1$\} \\  \hdashline
12 &Mg&\{ $1$, $2$\}&\{$2$\}& \{$2$\}& \{$2$\} \\  \hdashline
13 &Al&\{ $1$, $2$, $3$\}&\{$3$\}& \{$3$\}& \{$3$\} \\  \hdashline
14 &Si&\{ $-4$, $-3$, $-2$, $-1$, $1$, $2$, $3$, $4$\}&\{$-4$, $4$\}&\{$-4$, $4$\}&\{$-4$, $4$\} \\  \hdashline
15 &P&\{ $-3$, $-2$, $-1$, $1$, $2$, $3$, $4$, $5$\}&\{$-3$, $-2$, $-1$, $3$, $4$, $5$\}&\{$-3$, $3$, $5$\}&\{$-3$, $3$, $5$\} \\  \hdashline
16 &S&\{ $-2$, $-1$, $1$, $2$, $3$, $4$, $5$, $6$\}&\{$-2$, $-1$, $2$, $4$, $6$\}&\{$-2$, $2$, $4$, $6$\}&\{$-2$, $2$, $4$, $6$\} \\  \hdashline
17 &Cl&\{ $-1$, $1$, $2$, $3$, $4$, $5$, $6$, $7$\}&\{$-1$\}&\{$-1$, $1$, $3$, $5$, $7$\}&\{$-1$\} \\  \hdashline
18 &Ar& \{ \}& \{ \}& \{ \}& \{ \} \\  \hdashline
19 &K&\{ $-1$, $1$\}&\{$1$\}& \{$1$\}& \{$1$\} \\  \hdashline
20 &Ca&\{ $1$, $2$\}&\{$2$\}& \{$2$\}& \{$2$\} \\ 
%21 &Sc&\{ $1$, $2$, $3$\}&\{$2$, $3$\}&\{$3$\}& \{$3$\} \\  \hdashline
%22 &Ti&\{ $-1$, $1$, $2$, $3$, $4$\}&\{$2$, $3$, $4$\}&\{$4$\}& \{$4$\} \\  \hdashline
%23 &V&\{ $-1$, $1$, $2$, $3$, $4$, $5$\}&\{$2$, $3$, $4$, $5$\}&\{$5$\}& \{$5$\} \\ \hdashline
\hline%\bottomrule
\end{tabular}
\label{table__ox_state1}
\end{table}

\begin{table}[t]
\caption{List of oxidation numbers from Scandium (Sc) to Ytterbium (Yb).}
\centering
\begin{tabular}{wl{0.2cm}wl{0.2cm}wc{4.3cm}ccc}
\hline
Z&Elm&SMACT&icsd&common&Ours\\ \hline
21 &Sc&\{ $1$, $2$, $3$\}&\{$2$, $3$\}&\{$3$\}& \{$3$\} \\  \hdashline
22 &Ti&\{ $-1$, $1$, $2$, $3$, $4$\}&\{$2$, $3$, $4$\}&\{$4$\}& \{$4$\} \\  \hdashline
23 &V&\{ $-1$, $1$, $2$, $3$, $4$, $5$\}&\{$2$, $3$, $4$, $5$\}&\{$5$\}& \{$5$\} \\ \hdashline
24 &Cr&\{ $-2$, $-1$, $1$, $2$, $3$, $4$, $5$, $6$\}&\{$2$, $3$, $4$, $5$, $6$\}&\{$3$, $6$\}&\{$3$, $6$\} \\  \hdashline
25 &Mn&\{ $-3$, $-2$, $-1$, $1$, $2$, $3$, $4$, $5$, $6$, $7$\}&\{$2$, $3$, $4$, $7$\}&\{$2$, $4$, $7$\}&\{$2$, $4$, $7$\} \\  \hdashline
26 &Fe&\{ $-2$, $-1$, $1$, $2$, $3$, $4$, $5$, $6$\}&\{$2$, $3$\}&\{$2$, $3$\}&\{$2$, $3$\} \\  \hdashline
27 &Co&\{ $-1$, $1$, $2$, $3$, $4$, $5$\}&\{$1$, $2$, $3$, $4$\}&\{$2$, $3$\}&\{$2$, $3$\} \\  \hdashline
28 &Ni&\{ $-1$, $1$, $2$, $3$, $4$\}&\{$1$, $2$, $3$, $4$\}&\{$2$\}& \{$2$\} \\  \hdashline
29 &Cu&\{ $1$, $2$, $3$, $4$\}&\{$1$, $2$, $3$\}&\{$2$\}& \{$2$\} \\  \hdashline
30 &Zn&\{ $1$, $2$\}&\{$2$\}& \{$2$\}& \{$2$\} \\  \hdashline
31 &Ga&\{ $1$, $2$, $3$\}&\{$2$, $3$\}&\{$3$\}& \{$3$\} \\  \hdashline
32 &Ge&\{ $-4$, $-3$, $-2$, $-1$, $1$, $2$, $3$, $4$\}&\{$2$, $3$, $4$\}&\{$-4$, $2$, $4$\}&\{$2$, $4$\} \\  \hdashline
33 &As&\{ $-3$, $1$, $2$, $3$, $5$\}&\{$-3$, $-2$, $-1$, $2$, $3$, $5$\}&\{$-3$, $3$, $5$\}&\{$-3$, $3$, $5$\} \\  \hdashline
34 &Se&\{ $-2$, $1$, $2$, $4$, $6$\}&\{$-2$, $-1$, $4$, $6$\}&\{$-2$, $2$, $4$, $6$\}&\{$-2$, $4$, $6$\} \\  \hdashline
35 &Br&\{ $-1$, $1$, $2$, $3$, $4$, $5$, $7$\}&\{$-1$, $5$\}&\{$-1$, $1$, $3$, $5$, $7$\}&\{$-1$, $5$\} \\  \hdashline
36 &Kr& \{$2$\}& \{ \}& \{ \}& \{ \} \\  \hdashline
37 &Rb&\{ $-1$, $1$\}&\{$1$\}& \{$1$\}& \{$1$\} \\  \hdashline
38 &Sr&\{ $1$, $2$\}&\{$2$\}& \{$2$\}& \{$2$\} \\  \hdashline
39 &Y&\{ $1$, $2$, $3$\}&\{$3$\}& \{$3$\}& \{$3$\} \\  \hdashline
40 &Zr&\{ $1$, $2$, $3$, $4$\}&\{$2$, $3$, $4$\}&\{$4$\}& \{$4$\} \\  \hdashline
41 &Nb&\{ $-1$, $1$, $2$, $3$, $4$, $5$\}&\{$2$, $3$, $4$, $5$\}&\{$5$\}& \{$5$\} \\  \hdashline
42 &Mo&\{ $-2$, $-1$, $1$, $2$, $3$, $4$, $5$, $6$\}&\{$2$, $3$, $4$, $5$, $6$\}&\{$4$, $6$\}&\{$4$, $6$\} \\  \hdashline
43 &Tc&\{ $-3$, $-1$, $1$, $2$, $3$, $4$, $5$, $6$, $7$\}&\{ \}&\{$4$, $7$\}&\{ \} \\  \hdashline
44 &Ru&\{ $-2$, $1$, $2$, $3$, $4$, $5$, $6$, $7$, $8$\}&\{$2$, $3$, $4$, $5$, $6$\}&\{$3$, $4$\}&\{$3$, $4$\} \\  \hdashline
45 &Rh&\{ $-1$, $1$, $2$, $3$, $4$, $5$, $6$\}&\{$3$, $4$\}&\{$3$\}& \{$3$\} \\  \hdashline
46 &Pd&\{ $1$, $2$, $4$, $6$\}&\{$2$, $4$\}&\{$2$, $4$\}&\{$2$, $4$\} \\  \hdashline
47 &Ag&\{ $1$, $2$, $3$, $4$\}&\{$1$, $2$, $3$\}&\{$1$\}& \{$1$\} \\  \hdashline
48 &Cd&\{ $1$, $2$\}&\{$2$\}& \{$2$\}& \{$2$\} \\  \hdashline
49 &In&\{ $1$, $2$, $3$\}&\{$1$, $2$, $3$\}&\{$3$\}& \{$3$\} \\  \hdashline
50 &Sn&\{ $-4$, $2$, $4$\}&\{$2$, $3$, $4$\}&\{$-4$, $2$, $4$\}&\{$2$, $4$\} \\  \hdashline
51 &Sb&\{ $-3$, $3$, $5$\}&\{$-3$, $-2$, $-1$, $3$, $5$\}&\{$-3$, $3$, $5$\}&\{$-3$, $3$, $5$\} \\  \hdashline
52 &Te&\{ $-2$, $2$, $4$, $5$, $6$\}&\{$-2$, $-1$, $4$, $6$\}&\{$-2$, $2$, $4$, $6$\}&\{$-2$, $4$, $6$\} \\  \hdashline
53 &I&\{ $-1$, $1$, $3$, $4$, $5$, $7$\}&\{$-1$, $5$\}&\{$-1$, $1$, $3$, $5$, $7$\}&\{$-1$, $5$\} \\  \hdashline
54 &Xe&\{ $1$, $2$, $4$, $6$, $8$\}&\{ \}& \{ \}& \{ \} \\  \hdashline
55 &Cs&\{ $-1$, $1$\}&\{$1$\}& \{$1$\}& \{$1$\} \\  \hdashline 
56 &Ba& \{$2$\}& \{$2$\}& \{$2$\}& \{$2$\} \\  \hdashline
57 &La&\{ $2$, $3$\}&\{$2$, $3$\}&\{$3$\}& \{$3$\} \\  \hdashline
58 &Ce&\{ $2$, $3$, $4$\}&\{$3$, $4$\}&\{$3$, $4$\}&\{$3$, $4$\} \\  \hdashline
59 &Pr&\{ $2$, $3$, $4$\}&\{$3$, $4$\}&\{$3$\}& \{$3$\} \\  \hdashline
60 &Nd&\{ $2$, $3$, $4$\}&\{$2$, $3$\}&\{$3$\}& \{$3$\} \\  \hdashline
61 &Pm&\{ $2$, $3$\}&\{ \}& \{$3$\}& \{ \} \\  \hdashline
62 &Sm&\{ $2$, $3$\}&\{$2$, $3$\}&\{$3$\}& \{$3$\} \\  \hdashline
63 &Eu&\{ $2$, $3$\}&\{$2$, $3$\}&\{$2$, $3$\}&\{$2$, $3$\} \\  \hdashline
64 &Gd&\{ $1$, $2$, $3$\}&\{$3$\}& \{$3$\}& \{$3$\} \\  \hdashline 
65 &Tb&\{ $1$, $2$, $3$, $4$\}&\{$3$, $4$\}&\{$3$\}& \{$3$\} \\  \hdashline
66 &Dy&\{ $2$, $3$, $4$\}&\{$3$\}& \{$3$\}& \{$3$\} \\  \hdashline
67 &Ho&\{ $2$, $3$\}&\{$3$\}& \{$3$\}& \{$3$\} \\  \hdashline
68 &Er&\{ $2$, $3$\}&\{$3$\}& \{$3$\}& \{$3$\} \\  \hdashline
69 &Tm&\{ $2$, $3$\}&\{$3$\}& \{$3$\}& \{$3$\} \\  \hdashline
70 &Yb&\{ $2$, $3$\}&\{$2$, $3$\}&\{$3$\}& \{$3$\} \\  
\hline%\bottomrule
\end{tabular}
\label{table__ox_state2}
\end{table}

\clearpage 

\begin{table}[t]
\caption{List of oxidation numbers from Lutetium (Lu) to Californium (Cf)}
\centering
\begin{tabular}{wl{0.2cm}wl{0.2cm}wc{4.3cm}ccc}
\hline
Z&Elm&smact&icsd&common&Ours\\ \hline
71 &Lu& \{$3$\}& \{$3$\}& \{$3$\}& \{$3$\} \\  \hdashline
72 &Hf&\{ $2$, $3$, $4$\}&\{$4$\}& \{$4$\}& \{$4$\} \\  \hdashline
73 &Ta&\{ $-1$, $2$, $3$, $4$, $5$\}&\{$3$, $4$, $5$\}&\{$5$\}& \{$5$\} \\  \hdashline
74 &W&\{ $-2$, $-1$, $1$, $2$, $3$, $4$, $5$, $6$\}&\{$2$, $3$, $4$, $5$, $6$\}&\{$4$, $6$\}&\{$4$, $6$\} \\  \hdashline
75 &Re&\{ $-3$, $-1$, $1$, $2$, $3$, $4$, $5$, $6$, $7$\}&\{$3$, $4$, $5$, $6$, $7$\}&\{$4$\}& \{$4$\} \\  \hdashline
76 &Os&\{ $-2$, $-1$, $1$, $2$, $3$, $4$, $5$, $6$, $7$, $8$\}&\{ \}& \{$4$\}& \{ \} \\  \hdashline
77 &Ir&\{ $-3$, $-1$, $1$, $2$, $3$, $4$, $5$, $6$, $7$, $8$\}&\{$3$, $4$, $5$\}&\{$3$, $4$\}&\{$3$, $4$\} \\  \hdashline
78 &Pt&\{ $-2$, $-1$, $1$, $2$, $3$, $4$, $5$, $6$\}&\{ \}&\{$2$, $4$\}&\{ \} \\  \hdashline
79 &Au&\{ $-1$, $1$, $2$, $3$, $5$\}&\{ \}& \{$3$\}& \{ \} \\  \hdashline
80 &Hg&\{ $1$, $2$, $4$\}&\{$1$, $2$\}&\{$1$, $2$\}&\{$1$, $2$\} \\  \hdashline
81 &Tl&\{ $-1$, $1$, $3$\}&\{$1$, $3$\}&\{$1$, $3$\}&\{$1$, $3$\} \\  \hdashline
82 &Pb&\{ $-4$, $2$, $4$\}&\{$2$, $4$\}&\{$2$, $4$\}&\{$2$, $4$\} \\  \hdashline
83 &Bi&\{ $-3$, $1$, $3$, $5$, $7$\}&\{$1$, $2$, $3$, $5$\}&\{$3$\}& \{$3$\} \\  \hdashline
84 &Po&\{ $-2$, $2$, $4$, $5$, $6$\}&\{ \}&\{$-2$, $2$, $4$\}&\{ \} \\  \hdashline
85 &At&\{ $-1$, $1$, $3$, $5$, $7$\}&\{ \}&\{$-1$, $1$\}&\{ \} \\  \hdashline
86 &Rn&\{ $2$, $6$\}&\{ \}& \{ \}& \{ \} \\ 
87 &Fr& \{$1$\}& \{ \}& \{$1$\}& \{ \} \\  \hdashline
88 &Ra& \{$2$\}& \{ \}& \{$2$\}& \{ \} \\  \hdashline
89 &Ac&\{ $2$, $3$\}&\{ \}& \{$3$\}& \{ \} \\  \hdashline
90 &Th&\{ $2$, $3$, $4$\}&\{$4$\}& \{$4$\}& \{$4$\} \\  \hdashline
91 &Pa&\{ $2$, $3$, $4$, $5$\}&\{ \}& \{$5$\}& \{ \} \\  \hdashline
92 &U&\{ $2$, $3$, $4$, $5$, $6$\}&\{$3$, $4$, $5$, $6$\}&\{$6$\}& \{$6$\} \\  \hdashline
93 &Np&\{ $3$, $4$, $5$, $6$, $7$\}&\{ \}& \{$5$\}& \{ \} \\  \hdashline
94 &Pu&\{ $2$, $3$, $4$, $5$, $6$, $7$, $8$\}&\{ \}& \{$4$\}& \{ \} \\  \hdashline
95 &Am&\{ $2$, $3$, $4$, $5$, $6$, $7$\}&\{ \}& \{$3$\}& \{ \} \\  \hdashline
96 &Cm&\{ $2$, $3$, $4$, $6$, $8$\}&\{ \}& \{$3$\}& \{ \} \\  \hdashline
97 &Bk&\{ $2$, $3$, $4$\}&\{ \}& \{$3$\}& \{ \} \\  \hdashline
98 &Cf&\{ $2$, $3$, $4$\}&\{ \}& \{$3$\}& \{ \} \\  
\hline%\bottomrule
\end{tabular}
\label{table__ox_state3}
\end{table}

\clearpage

\section{Additional Experimental Results}

\subsection{Optimization runtime for SMOACS} \label{section__apx_run_time}

We report the approximate runtime of SMOACS optimization in different experimental settings. In the experiments presented in Section~\ref{section__init_experiment} and the unit-cell perovskite optimization in Section~\ref{sec__perov_experiments}, SMOACS with Crystalformer optimizes 256 samples in approximately two minutes. In the large-scale setting of $3\times3\times3$-size perovskite optimization in Section~\ref{sec__perov_experiments}, optimization of 32 samples takes approximately 15 minutes with ALIGNN and 35 minutes with Crystalformer. In Section~\ref{sec__comparison_with_mattergen}, 128 samples were optimized in less than five minutes. In Section~\ref{sec__half_metal}, 150 samples were optimized in approximately 15 minutes.

% half-metal では、150サンプルの最適化に15分程度
These runtimes were measured using a single NVIDIA A100 GPU.

\subsection{Detailed results of initialization test} \label{appendix__details_of_init_test}

Table~\ref{table__summary_initialization_mutation_analysis_all} presents the full results corresponding to Table~\ref{table__summary_initialization_mutation_analysis}, with standard deviations included to indicate the variability across multiple runs. The notations and experimental settings remain consistent with those used in the main text.

We applied simulated annealing (SA) after the SMOACS optimization step to evaluate whether perturbations to the candidate structure led to improved loss values. Zero-mean Gaussian noise is independently added to the lattice constants, atomic coordinates, and atomic distribution, with the standard deviation controlling the noise magnitude. We tuned the standard deviation values over \{0.1, 0.01, 0.001, 0.0001, 0.00001\}, initial temperatures over \{0.1, 1.0, 10.0, 100.0\}, and exponential decay rates of the temperature schedule over \{0.99, 0.9, 0.8, 0.7, 0.5, 0.25\}. Based on empirical performance, we selected the final configuration as standard deviation = 0.1, initial temperature = 0.1, and decay rate = 0.9.

\begin{table}[ht]
\centering
\caption{Results with standard deviations added to Table~\ref{table__summary_initialization_mutation_analysis}. Notations follow Table~\ref{table__summary_initialization_mutation_analysis}}
\label{table__summary_initialization_mutation_analysis_all}
\begin{tabular}{wl{2.1cm}wl{1.4cm}wl{0.9cm}wc{1.1cm}wc{1.0cm}wc{1.0cm}wc{1.0cm}wc{1.0cm}wc{1.0cm}}
\hline \toprule
    Opt. Method & Ox. Restr. & \makecell[l]{Struct.\\Init.}  &\begin{tabular}{c}Success\\rate\end{tabular} & (i) $E_g$  &  (ii) $E_f$&(iii) Neu. &  Uniq. &  Novel \\ \midrule
    Grad. & None & Rand. & \begin{tabular}{c}$0.00 $\\ $\pm0.00$\end{tabular} & \begin{tabular}{c}$0.30 $\\ $\pm0.02$\end{tabular} & \begin{tabular}{c}$0.06 $\\ $\pm0.02$\end{tabular} & \begin{tabular}{c}$0.01 $\\ $\pm0.00$\end{tabular} & N/A & N/A \\
    Grad. + SA & None & Rand. & \begin{tabular}{c}$0.00 $\\ $\pm0.00$\end{tabular} & \begin{tabular}{c}$0.47 $\\ $\pm0.03$\end{tabular} & \begin{tabular}{c}$0.12 $\\ $\pm0.01$\end{tabular} & \begin{tabular}{c}$0.01 $\\ $\pm0.01$\end{tabular} & N/A & N/A \\
    Grad. & None & Templ. & \begin{tabular}{c}$0.07 $\\ $\pm0.02$\end{tabular} & \begin{tabular}{c}$0.65 $\\ $\pm0.01$\end{tabular} & \begin{tabular}{c}$0.83 $\\ $\pm0.01$\end{tabular} & \begin{tabular}{c}$0.11 $\\ $\pm0.03$\end{tabular} & \begin{tabular}{c}$0.97 $\\ $\pm0.06$\end{tabular} & \begin{tabular}{c}$0.93 $\\ $\pm0.05$\end{tabular} \\
    Grad. & Ox. Mask & Templ. & \begin{tabular}{c}$0.42 $\\ $\pm0.03$\end{tabular} & \begin{tabular}{c}$0.58 $\\ $\pm0.04$\end{tabular} & \begin{tabular}{c}$0.62 $\\ $\pm0.03$\end{tabular} & \begin{tabular}{c}$\underline{\mathbf{1.00 }}$\\ $\underline{\mathbf{ \pm 0.00}}$\end{tabular} & \begin{tabular}{c}$1.00 $\\ $\pm0.01$\end{tabular} & \begin{tabular}{c}$0.96 $\\ $\pm0.02$\end{tabular} \\
    Grad. + SA & Ox. Mask & Templ. & \begin{tabular}{c}$0.47 $\\ $\pm0.03$\end{tabular} & \begin{tabular}{c}$0.52 $\\ $\pm0.03$\end{tabular} & \begin{tabular}{c}$0.75 $\\ $\pm0.01$\end{tabular} & \begin{tabular}{c}$\underline{\mathbf{1.00 }}$\\ $\underline{\mathbf{ \pm 0.00}}$\end{tabular} & \begin{tabular}{c}$0.99 $\\ $\pm0.01$\end{tabular} & \begin{tabular}{c}$0.88 $\\ $\pm0.02$\end{tabular} \\
    Grad. + Mut. & Ox. Mask & Templ. & \begin{tabular}{c}$0.64 $\\ $\pm0.02$\end{tabular} & \begin{tabular}{c}$0.67 $\\ $\pm0.02$\end{tabular} & \begin{tabular}{c}$0.93 $\\ $\pm0.01$\end{tabular} & \begin{tabular}{c}$\underline{\mathbf{1.00 }}$\\ $\underline{\mathbf{ \pm 0.00}}$\end{tabular} & \begin{tabular}{c}$0.99 $\\ $\pm0.01$\end{tabular} & \begin{tabular}{c}$0.92 $\\ $\pm0.01$\end{tabular} \\
    Grad. + SA/Mut. & Ox. Mask & Templ. & \begin{tabular}{c}$\underline{\mathbf{0.67 }}$\\ $\underline{\mathbf{ \pm 0.03}}$\end{tabular} & \begin{tabular}{c}$\underline{\mathbf{0.70 }}$\\ $\underline{\mathbf{ \pm 0.04}}$\end{tabular} & \begin{tabular}{c}$\underline{\mathbf{0.94 }}$\\ $\underline{\mathbf{ \pm 0.02}}$\end{tabular} & \begin{tabular}{c}$\underline{\mathbf{1.00 }}$\\ $\underline{\mathbf{ \pm 0.00}}$\end{tabular} & \begin{tabular}{c}$1.00 $\\ $\pm0.00$\end{tabular} & \begin{tabular}{c}$0.97 $\\ $\pm0.01$\end{tabular} \\
\bottomrule
\end{tabular}
\end{table}

\subsection{Targeting exceptional properties} \label{appendix__section__rare_experiment}

Exceptional properties---such as high-$T_c$ superconductivity and wide band gaps---are critical in industry but difficult to achieve using diffusion models~\cite{fujii2025straightforward}. We assessed SMOACS's ability by optimizing for wide band gaps using MEGNet predictors and high $T_c$ using models trained on the JARVIS-Superconductor dataset~\cite{choudhary2020joint}, leveraging pretrained ALIGNN or Crystalformer without retraining. We use the same loss and targets as in Section~\ref{section__init_experiment} for band gap. For superconductivity, let $\hat{y}_{T_c}$ denote the predicted superconducting transition temperature. We minimize $L_{T_c} = -\hat{y}_{T_c}$.

For comparison, we followed the official implementation, which trains a generative model and predictor and then optimizes latent $z$ using the same loss, starting from structures sampled from each dataset. This loss guides $z$ toward exceptional regions. CDVAE performs optimization in latent space, generates crystals through denoising, and then re-encodes the generated crystals into latent space to evaluate properties (see SI section~\ref{appendix__implement_details} for details). 

%For comparison, we follow the official CDVAE implementation, which jointly trains a generative model and predictor, then optimizes latent variables $z$ using the same loss  starting from crystals from each dataset. This loss implicitly guides $z$ toward exceptional regions, conceptually similar to a minority-guided loss~\cite{um2023don}. CDVAE performs optimization in latent space, generates crystals through denoising, and then re-encodes the generated crystals into latent space to evaluate properties. (see SI section~\ref{appendix__implement_details} for details). 

Tables~\ref{table__rare_property_tc} and \ref{table__rare_property_bg} show success rates for optimizing $T_c$ and band gap. CDVAE performs moderately when optimizing latent variables, but its performance degrades after generating crystals due to property shifts. However, its success rates decline significantly for more exceptional targets, indicating that such guidance strategies may offer limited benefit in materials design~\cite{fujii2025straightforward} (for a detailed discussion, see SI section~\ref{section__cdvae_analysis}). In contrast, SMOACS remains effective even for exceptional properties.

\begin{table}
\centering
\caption{Success rates for achieving predicted $T_c$ above various thresholds. "CDVAE" shows $T_c$ after structure generation from optimized $z$. Percentiles are based on the JARVIS Superconductor dataset.}
\label{table__rare_property_tc}
\begin{tabular}{lcccc}
\midrule
    \begin{tabular}{c}$T_c$ \\ (Percentile)\end{tabular} & \begin{tabular}{c}$>$3.0 K\\($>$60.3\%)\end{tabular} & \begin{tabular}{c}$>$5.0 K\\($>$73.3\%)\end{tabular} & \begin{tabular}{c}$>$10.0 K\\($>$90.1\%)\end{tabular} & \begin{tabular}{c}$>$15.0 K\\($>$96.3\%)\end{tabular} \\
\midrule    
    CDVAE & $0.43\pm0.07$ & $0.22\pm0.07$ & $0.03\pm0.03$ & $0.02\pm0.02$ \\
    SMOACS & $\underline{\mathbf{1.00\pm0.00}}$ & $\underline{\mathbf{0.92\pm0.03}}$ & $\underline{\mathbf{0.54\pm0.02}}$ & $\underline{\mathbf{0.25\pm0.02}}$ \\
\bottomrule
\end{tabular}
\end{table}

\begin{table}
\centering
\caption{Success rates for targeted band gap values. S(Cry) and S(ALI) denote SMOACS with Crystalformer and ALIGNN. Other labels follow Table~\ref{table__rare_property_tc}. Percentiles follow the MEGNet dataset.}
\label{table__rare_property_bg}
\begin{tabular}{l|cccc}
\midrule
     \begin{tabular}{c}Target $E_g$ (eV) \\(Percentile)\end{tabular} & \begin{tabular}{c}$3.5 \pm 0.2$\\(87.6\%)\end{tabular} & \begin{tabular}{c}$4.0 \pm 0.2$\\(91.6\%)\end{tabular} & \begin{tabular}{c}$4.5 \pm 0.2$\\(94.4\%)\end{tabular} & \begin{tabular}{c}$5.0 \pm 0.2$\\(96.5\%)\end{tabular} \\
\toprule    
    CDVAE & $0.03\pm0.01$ & $0.01\pm0.01$ & $0.01\pm0.01$ & $0.02\pm0.01$ \\
    S(ALI) & $\underline{\mathbf{0.70\pm0.02}}$ & $\underline{\mathbf{0.67\pm0.04}}$ & $0.64\pm0.02$ & $0.61\pm0.05$ \\
    S(Cry) & $0.65\pm0.03$ & $0.67\pm0.02$ & $\underline{\mathbf{0.65\pm0.03}}$ & $\underline{\mathbf{0.67\pm0.01}}$ \\
\bottomrule
\end{tabular}
\end{table}

\subsection{Discussion: why CDVAE struggles with property optimization} \label{section__cdvae_analysis}

Tables~\ref{table__rare_property_tc__apx} and \ref{table__rare_property_bg__apx} show success rates for optimizing $T_c$ and band gap. CDVAE performs moderately when optimizing latent variables (CDVAE (opt. $z$) in these tables), but its performance degrades after generating crystals due to property shifts. CDVAE adopts a loss function similar in spirit to the minority-guided loss. However, its success rates decline significantly for more exceptional targets.

We attribute the limited performance of CDVAE in optimizing for exceptional properties to two main factors:

\textbf{(1) Property shifts caused by volume change from reconstruction:} 
CDVAE’s two-step process---latent optimization followed by denoising-based reconstruction---introduces instability in properties. In our MEGNet dataset experiments, reconstructed crystals show an average 20\% volume change. Such volume changes are well known to significantly impact the band gap. Even when using the much larger MP20 dataset (which contains over 1.6 million entries and is nearly 25 times the size of the MEGNet dataset) as in the original CDVAE paper~\cite{xiecrystal}, 10\% volume shifts remain. This indicates that the issue is intrinsic to the challenge of training generative models on datasets that include crystals of various sizes, rather than being solely due to dataset size.

\textbf{(2) Reversion during denoising:}
Our analysis of the latent-to-crystal property differences ("Opt-to-Cryst Diff" in Tables~\ref{table__rare_property_tc__apx} and \ref{table__rare_property_bg__apx}) shows a clear trend: generated crystals systematically shift away from minority regions in property space. If volume changes were random, however, one would expect near-random property shifts. This aligns with Um \textit{et al.}~\cite{um2023don}, who show that diffusion models revert towards dominant regions during denoising, hindering minority sample generation. To address this, they propose property-guided optimization with a classifier-based predictor. Our method similarly uses property guidance but applies it to crystal structure optimization with regression-based predictors. Notably, Fujii \textit{et al.}~\cite{fujii2025straightforward} implemented property guidance iteratively during denoising steps, yet still failed to reliably steer generation toward minority regions.

In contrast, SMOACS directly optimizes crystal structures without relying on latent space or property-guided denoising, avoiding instability and achieving higher success rates in materials design tasks.

\begin{table}[hb]
\centering
\caption{Success rates (SC rate) for achieving a predicted $T_c$ above various thresholds. "SC rate: CDVAE (opt. $z$)" shows the success rate achieved in the latent space optimization (i.e., optimized $z$). "SC rate: CDVAE" shows the success rate after generating crystal structures from the optimized latent variables $z$. "Opt-to-Cryst Diff" represents the difference in $T_c$ between the optimized latent variables $z$ and their decoded crystal structures (a negative value indicates a decrease after decoding). Percentiles are calculated based on the JARVIS Superconductor dataset.}
\label{table__rare_property_tc__apx}
\begin{tabular}{lcccc}
\toprule
\begin{tabular}{c}$T_c$ \\ (Percentile)\end{tabular} & \begin{tabular}{c}3.0 K\\(60.3\%)\end{tabular} & \begin{tabular}{c}5.0 K\\(73.3\%)\end{tabular} & \begin{tabular}{c}10.0 K\\(90.1\%)\end{tabular} & \begin{tabular}{c}20.0 K\\(98.5\%)\end{tabular} \\
\midrule    
    SC rate: CDVAE (opt. $z$) & $0.71\pm0.04$ & $0.71\pm0.04$ & $0.71\pm0.04$ & $0.71\pm0.04$ \\
    SC rate: CDVAE & $0.48\pm0.02$ & $0.23\pm0.09$ & $0.03\pm0.04$ & $0.00\pm0.00$ \\
   \hdashline Opt-to-Cryst Diff (K) & $-0.28\pm1.79$ & $-1.22\pm2.08$ & $-3.43\pm4.08$ & $-13.04\pm4.61$ \\
\bottomrule
\end{tabular}
\end{table}

\begin{table}[hb]
\centering
\caption{Success rates for target band gaps. Notations follow Table~\ref{table__rare_property_tc__apx}. Percentiles follow the MEGNet dataset.}
\label{table__rare_property_bg__apx}
\begin{tabular}{lcccc}
\toprule
 \begin{tabular}{c}Target $E_g$ (eV) \\(Percentile)\end{tabular} & \begin{tabular}{c}$2.0 \pm 0.2$\\(69.3\%)\end{tabular} & \begin{tabular}{c}$3.0 \pm 0.2$\\(82.4\%)\end{tabular} & \begin{tabular}{c}$4.0 \pm 0.2$\\(91.6\%)\end{tabular} & \begin{tabular}{c}$5.0 \pm 0.2$\\(96.5\%)\end{tabular} \\
\midrule    
    SC rate: CDVAE (opt. $z$) & $0.36\pm0.02$ & $0.37\pm0.01$ & $0.34\pm0.01$ & $0.30\pm0.01$ \\
    SC rate: CDVAE & $0.05\pm0.02$ & $0.03\pm0.01$ & $0.01\pm0.01$ & $0.02\pm0.01$ \\
   \hdashline Opt-to-Cryst Diff (eV) & $-0.40\pm1.49$ & $-1.04\pm1.56$ & $-1.74\pm1.65$ & $-2.40\pm1.76$ \\
\bottomrule
\end{tabular}
\end{table}

%SMOACS achieved significantly lower scores with ALIGNN compared to Crystalformer. We attribute this to changes in the loss landscape resulting from updates to the graph structure, as discussed in SI section~\ref{appendix__details_of_random_crystal_bg_opt}.

\clearpage

\subsection{A possible application: identifying the most stable crystal structure}
\label{identifying_stable_structure}

Our method can optimize energy while specifying the base crystal structure. This property may allow for identifying crystal structures based either solely on the chemical formula or on a combination of the chemical formula and physical properties. This is a Crystal Structure Prediction (CSP) task~\cite{ryan2018crystal}. To test this possibility, we examined whether the crystal structure of metallic silicon with a zero band gap could be identified. We first extracted one-atom structures from the MEGNet dataset and used them as initial structures. The atomic distribution was fixed with a one-hot vector indicating silicon, and only the lattice constants were optimized. The target properties for optimization were a zero band gap and formation energy minimization. We chose silicon structures from the MEGNet dataset with a band gap of 0 eV as the reference and compared these with the optimized structures that exhibited the lowest formation energy. Consequently, we identified structures close to the reference among those optimized for the lowest formation energy.

The results are shown in Table~\ref{table__poc_identification}. The reference material mp-34 is close to optimized candidate 2. Similarly, mp-1014212 is close to candidates 4--12.

\begin{table}[hp]
\caption{Reference Si materials (band gap 0 eV) and optimized candidates.}
\centering
\begin{tabular}{cccc}
\hline
Materials & $a$,$b$,$c$ (\r{A})& $\alpha$,$\beta$,$\gamma$ (°) & \begin{tabular}{c} predicted formation\\energy (eV/atom) \end{tabular}  \\
\hline
%(Ref) mp-27 $ & $2.73,2.73,2.73$ & $ 60.0, 60.0, 60.0$ & -\\ \hdashline
(Ref) mp-34  & $2.64,2.64,2.47$ & $ 90.0, 90.0,120.0$ & -\\ \hdashline
(Ref) mp-1014212 & $2.66,2.66,2.66$ & $109.5,109.5,109.5$ & -\\
\hline
candidate-1 & $2.67,2.67,2.94$ & $124.0,124.0,97.9$ & $-0.367$\\ \hdashline
candidate-2 & $2.50,2.50,2.27$ & $89.9,89.9,134.0$ & $-0.359$\\ \hdashline
candidate-3 & $2.76,2.76,2.76$ & $115.3,115.3,115.3$ & $-0.326$\\ \hdashline
candidate-4 & $2.72,2.72,2.72$ & $115.0,115.0,115.0$ & $-0.310$\\ \hdashline
candidate-5 & $2.71,2.71,2.71$ & $114.9,114.9,114.9$ & $-0.310$\\ \hdashline
candidate-6 & $2.71,2.71,2.71$ & $114.9,114.9,114.9$ & $-0.310$\\ \hdashline
candidate-7 & $2.71,2.71,2.71$ & $114.9,114.9,114.9$ & $-0.308$\\ \hdashline
candidate-8 & $2.73,2.73,2.73$ & $115.0,115.0,115.0$ & $-0.308$\\ \hdashline
candidate-9 & $2.69,2.69,2.69$ & $114.9,114.9,114.9$ & $-0.308$\\ \hdashline
candidate-10 & $2.68,2.68,2.68$ & $114.8,114.8,114.8$ & $-0.304$\\ \hdashline
candidate-11 & $2.64,2.64,2.64$ & $114.3,114.3,114.3$ & $-0.266$\\ \hdashline
candidate-12 & $2.55,2.55,2.55$ & $113.9,113.9,113.9$ & $-0.263$\\ \hdashline

\hline
\end{tabular}
\label{table__poc_identification}
\end{table}

\subsection{Details in multi-property design of perovskite structures} \label{appendix__details_of_perov_crystal_bg_opt}

We evaluated whether the optimized structures approximated typical perovskite configurations.
Typical fractional coordinates of perovskite structures are as follows: $(0.5, 0.5, 0.5)$ at the A site, $(0.0, 0.0, 0.0)$ at the B site, and $(0.5, 0.0, 0.0), (0.0, 0.5, 0.0), (0.0, 0.0, 0.5)$ at the three X sites. We established criteria for the optimized x, y, and z coordinates to be within a deviation $\epsilon$ from these standard values. The perovskite structure $\mathrm{CaCu_3Ti_4O_{12}}$ exhibits a slightly distorted configuration, with the x-coordinate of the oxygen atoms deviating by approximately 10\% from their typical positions~\cite{ESBozin_2004}. To explore new structures, we set $\epsilon = 0.15$ for the unit cell and $\epsilon = 0.05$ for the $3\times3\times3$-size cell, allowing for a slightly greater distortion. We considered the optimized coordinates successful if the x, y, and z coordinates of each site fell within $\pm \epsilon$. Additionally, the angles between the crystal axes of typical perovskite structures are close to 90°. Therefore, angles between 85° and 95° were established as a criterion.

Using $t$ values from typical perovskite structures ($\mathrm{BaCeO_3}$: 0.857, $\mathrm{SrTiO_3}$: 0.910, and $\mathrm{BaTiO_3}$: 0.970), we established a tolerance-factor range of $0.8 \leq t \leq 1.0$ as the success criterion. The ionic radius of the X site was calculated as the average of the radii of the three X sites. We took the values for the ionic radii from pymatgen~\cite{ONG2013314}.

Due to the limited number of perovskite structure data points in the MEGNet dataset, we generated random perovskite structures as initial values for SMOACS and CDVAE. These structures have crystal axis angles $\alpha$, $\beta$, and $\gamma$ at 90° and axis lengths $a$, $b$, and $c$ randomly generated between 2 \r{A} and 10 \r{A}. Their initial fractional coordinates correspond to those typical of perovskite structures: $(0.5, 0.5, 0.5)$ for the A site, $(0.0, 0.0, 0.0)$ for the B site, and $(0.5, 0.0, 0.0)$, $(0.0, 0.5, 0.0)$, and $(0.0, 0.0, 0.5)$ for the three X sites. Similarly, TPE optimized perovskite structures by setting the crystal axis angles at 90° and optimizing the axis lengths $a, b, c$ between 2 \r{A} and 10 \r{A}. We also limited element species for each site in TPE. Specifically, the elements are restricted by oxidation numbers: $+1$ and $+2$ for site A, $+2$ and $+4$ for site B, and $-1$ and $-2$ for site X. For FTCP, we initially selected data points where the crystal axis angles were at 90°, and all sites conformed to the typical fractional coordinates of perovskite structures; these were then converted into latent variables. 

To ensure a fair comparison, CDVAE began optimization from the same initial structures as SMOACS in all experiments. We also confirmed that starting optimization from perovskite structures within the dataset did not significantly affect the results.

\begin{table}
\centering
\small
\caption{Experiments on optimizing various band gaps while preserving perovskite structures. The "SC rate" reflects the probability of simultaneously satisfying five criteria. Criteria (i)-(v) follow Table~\ref{table__perovskite_eval}. "Uniq." is the fraction of samples with unique element combinations among successful samples.}
\label{table__perovskite_eval_all}
\begin{tabular}{llwc{1.2cm}wc{1.0cm}wc{1.0cm}wc{1.0cm}wc{1.0cm}wc{1.0cm}wc{1.0cm}wc{1.0cm}wc{1.0cm}wc{1.0cm}}
\toprule
     target $E_g$ (eV) & Method & SC rate &(i)$E_g$ &(ii)$E_f$ &  (iii) $t$ &(iv)Neu. & (v)Prv. &  Uniq. \\
\midrule
    \multirow{5}{*}{$0.5 \pm 0.2$} & FTCP & \begin{tabular}{c}$0.00 $\\ $\pm0.00$\end{tabular} & \begin{tabular}{c}$0.00 $\\ $\pm0.00$\end{tabular} & \begin{tabular}{c}$\underline{\mathbf{1.00 }}$\\ $\underline{\mathbf{ \pm 0.00}}$\end{tabular} & \begin{tabular}{c}$0.23 $\\ $\pm0.05$\end{tabular} & \begin{tabular}{c}$0.99 $\\ $\pm0.00$\end{tabular} & \begin{tabular}{c}$0.23 $\\ $\pm0.05$\end{tabular} & N/A \\  \cdashline{2-9}
     & CDVAE & \begin{tabular}{c}$0.00 $\\ $\pm0.00$\end{tabular} & \begin{tabular}{c}$0.21 $\\ $\pm0.03$\end{tabular} & \begin{tabular}{c}$0.42 $\\ $\pm0.16$\end{tabular} & \begin{tabular}{c}$0.00 $\\ $\pm0.00$\end{tabular} & \begin{tabular}{c}$0.38 $\\ $\pm0.11$\end{tabular} & \begin{tabular}{c}$0.00 $\\ $\pm0.00$\end{tabular} & N/A \\  \cdashline{2-9}
     & TPE & \begin{tabular}{c}$0.09 $\\ $\pm0.01$\end{tabular} & \begin{tabular}{c}$\underline{\mathbf{1.00 }}$\\ $\underline{\mathbf{ \pm 0.00}}$\end{tabular} & \begin{tabular}{c}$0.46 $\\ $\pm0.05$\end{tabular} & \begin{tabular}{c}$0.30 $\\ $\pm0.01$\end{tabular} & \begin{tabular}{c}$0.95 $\\ $\pm0.01$\end{tabular} & \begin{tabular}{c}$\underline{\mathbf{1.00 }}$\\ $\underline{\mathbf{ \pm 0.00}}$\end{tabular} & \begin{tabular}{c}$1.00 $\\ $\pm0.00$\end{tabular} \\  \cdashline{2-9}
     & S(ALI) & \begin{tabular}{c}$0.20 $\\ $\pm0.02$\end{tabular} & \begin{tabular}{c}$0.71 $\\ $\pm0.02$\end{tabular} & \begin{tabular}{c}$0.46 $\\ $\pm0.03$\end{tabular} & \begin{tabular}{c}$\underline{\mathbf{0.44 }}$\\ $\underline{\mathbf{ \pm 0.03}}$\end{tabular} & \begin{tabular}{c}$\underline{\mathbf{1.00 }}$\\ $\underline{\mathbf{ \pm 0.00}}$\end{tabular} & \begin{tabular}{c}$\underline{\mathbf{1.00 }}$\\ $\underline{\mathbf{ \pm 0.00}}$\end{tabular} & \begin{tabular}{c}$0.54 $\\ $\pm0.07$\end{tabular} \\  \cdashline{2-9}
     & S(Cry) & \begin{tabular}{c}$\underline{\mathbf{0.23 }}$\\ $\underline{\mathbf{ \pm 0.03}}$\end{tabular} & \begin{tabular}{c}$0.62 $\\ $\pm0.01$\end{tabular} & \begin{tabular}{c}$0.86 $\\ $\pm0.03$\end{tabular} & \begin{tabular}{c}$0.37 $\\ $\pm0.04$\end{tabular} & \begin{tabular}{c}$\underline{\mathbf{1.00 }}$\\ $\underline{\mathbf{ \pm 0.00}}$\end{tabular} & \begin{tabular}{c}$\underline{\mathbf{1.00 }}$\\ $\underline{\mathbf{ \pm 0.00}}$\end{tabular} & \begin{tabular}{c}$0.90 $\\ $\pm0.03$\end{tabular} \\
\midrule    
    \multirow{5}{*}{$2.5 \pm 0.2$} & FTCP & \begin{tabular}{c}$0.00 $\\ $\pm0.00$\end{tabular} & \begin{tabular}{c}$0.00 $\\ $\pm0.00$\end{tabular} & \begin{tabular}{c}$\underline{\mathbf{1.00 }}$\\ $\underline{\mathbf{ \pm 0.00}}$\end{tabular} & \begin{tabular}{c}$0.24 $\\ $\pm0.04$\end{tabular} & \begin{tabular}{c}$0.99 $\\ $\pm0.01$\end{tabular} & \begin{tabular}{c}$0.24 $\\ $\pm0.04$\end{tabular} & N/A \\  \cdashline{2-9}
     & CDVAE & \begin{tabular}{c}$0.00 $\\ $\pm0.00$\end{tabular} & \begin{tabular}{c}$0.04 $\\ $\pm0.02$\end{tabular} & \begin{tabular}{c}$0.41 $\\ $\pm0.06$\end{tabular} & \begin{tabular}{c}$0.00 $\\ $\pm0.00$\end{tabular} & \begin{tabular}{c}$0.38 $\\ $\pm0.11$\end{tabular} & \begin{tabular}{c}$0.00 $\\ $\pm0.00$\end{tabular} & N/A \\  \cdashline{2-9}
     & TPE & \begin{tabular}{c}$0.06 $\\ $\pm0.01$\end{tabular} & \begin{tabular}{c}$\underline{\mathbf{1.00 }}$\\ $\underline{\mathbf{ \pm 0.00}}$\end{tabular} & \begin{tabular}{c}$0.53 $\\ $\pm0.05$\end{tabular} & \begin{tabular}{c}$0.22 $\\ $\pm0.04$\end{tabular} & \begin{tabular}{c}$0.70 $\\ $\pm0.05$\end{tabular} & \begin{tabular}{c}$\underline{\mathbf{1.00 }}$\\ $\underline{\mathbf{ \pm 0.00}}$\end{tabular} & \begin{tabular}{c}$1.00 $\\ $\pm0.00$\end{tabular} \\  \cdashline{2-9}
     & S(ALI) & \begin{tabular}{c}$\underline{\mathbf{0.39 }}$\\ $\underline{\mathbf{ \pm 0.04}}$\end{tabular} & \begin{tabular}{c}$0.63 $\\ $\pm0.05$\end{tabular} & \begin{tabular}{c}$0.96 $\\ $\pm0.01$\end{tabular} & \begin{tabular}{c}$\underline{\mathbf{0.64 }}$\\ $\underline{\mathbf{ \pm 0.02}}$\end{tabular} & \begin{tabular}{c}$\underline{\mathbf{1.00 }}$\\ $\underline{\mathbf{ \pm 0.00}}$\end{tabular} & \begin{tabular}{c}$\underline{\mathbf{1.00 }}$\\ $\underline{\mathbf{ \pm 0.00}}$\end{tabular} & \begin{tabular}{c}$0.51 $\\ $\pm0.04$\end{tabular} \\  \cdashline{2-9}
     & S(Cry) & \begin{tabular}{c}$0.24 $\\ $\pm0.01$\end{tabular} & \begin{tabular}{c}$0.56 $\\ $\pm0.01$\end{tabular} & \begin{tabular}{c}$0.88 $\\ $\pm0.02$\end{tabular} & \begin{tabular}{c}$0.46 $\\ $\pm0.01$\end{tabular} & \begin{tabular}{c}$\underline{\mathbf{1.00 }}$\\ $\underline{\mathbf{ \pm 0.00}}$\end{tabular} & \begin{tabular}{c}$\underline{\mathbf{1.00 }}$\\ $\underline{\mathbf{ \pm 0.00}}$\end{tabular} & \begin{tabular}{c}$0.93 $\\ $\pm0.05$\end{tabular} \\
\midrule    
    \multirow{5}{*}{$4.0 \pm 0.2$} & FTCP & \begin{tabular}{c}$0.00 $\\ $\pm0.00$\end{tabular} & \begin{tabular}{c}$0.00 $\\ $\pm0.00$\end{tabular} & \begin{tabular}{c}$\underline{\mathbf{1.00 }}$\\ $\underline{\mathbf{ \pm 0.00}}$\end{tabular} & \begin{tabular}{c}$0.22 $\\ $\pm0.05$\end{tabular} & \begin{tabular}{c}$0.99 $\\ $\pm0.01$\end{tabular} & \begin{tabular}{c}$0.22 $\\ $\pm0.05$\end{tabular} & N/A \\  \cdashline{2-9}
     & CDVAE & \begin{tabular}{c}$0.00 $\\ $\pm0.00$\end{tabular} & \begin{tabular}{c}$0.03 $\\ $\pm0.03$\end{tabular} & \begin{tabular}{c}$0.42 $\\ $\pm0.18$\end{tabular} & \begin{tabular}{c}$0.00 $\\ $\pm0.00$\end{tabular} & \begin{tabular}{c}$0.38 $\\ $\pm0.10$\end{tabular} & \begin{tabular}{c}$0.00 $\\ $\pm0.00$\end{tabular} & N/A \\  \cdashline{2-9}
     & TPE & \begin{tabular}{c}$0.04 $\\ $\pm0.02$\end{tabular} & \begin{tabular}{c}$\underline{\mathbf{0.96 }}$\\ $\underline{\mathbf{ \pm 0.02}}$\end{tabular} & \begin{tabular}{c}$0.43 $\\ $\pm0.06$\end{tabular} & \begin{tabular}{c}$0.20 $\\ $\pm0.01$\end{tabular} & \begin{tabular}{c}$0.40 $\\ $\pm0.07$\end{tabular} & \begin{tabular}{c}$\underline{\mathbf{1.00 }}$\\ $\underline{\mathbf{ \pm 0.00}}$\end{tabular} & \begin{tabular}{c}$0.96 $\\ $\pm0.08$\end{tabular} \\  \cdashline{2-9}
     & S(ALI) & \begin{tabular}{c}$\underline{\mathbf{0.36 }}$\\ $\underline{\mathbf{ \pm 0.03}}$\end{tabular} & \begin{tabular}{c}$0.60 $\\ $\pm0.04$\end{tabular} & \begin{tabular}{c}$0.99 $\\ $\pm0.01$\end{tabular} & \begin{tabular}{c}$\underline{\mathbf{0.62 }}$\\ $\underline{\mathbf{ \pm 0.02}}$\end{tabular} & \begin{tabular}{c}$\underline{\mathbf{1.00 }}$\\ $\underline{\mathbf{ \pm 0.00}}$\end{tabular} & \begin{tabular}{c}$\underline{\mathbf{1.00 }}$\\ $\underline{\mathbf{ \pm 0.00}}$\end{tabular} & \begin{tabular}{c}$0.31 $\\ $\pm0.02$\end{tabular} \\  \cdashline{2-9}
     & S(Cry) & \begin{tabular}{c}$0.29 $\\ $\pm0.01$\end{tabular} & \begin{tabular}{c}$0.62 $\\ $\pm0.04$\end{tabular} & \begin{tabular}{c}$0.98 $\\ $\pm0.02$\end{tabular} & \begin{tabular}{c}$0.46 $\\ $\pm0.04$\end{tabular} & \begin{tabular}{c}$\underline{\mathbf{1.00 }}$\\ $\underline{\mathbf{ \pm 0.00}}$\end{tabular} & \begin{tabular}{c}$\underline{\mathbf{1.00 }}$\\ $\underline{\mathbf{ \pm 0.00}}$\end{tabular} & \begin{tabular}{c}$0.70 $\\ $\pm0.02$\end{tabular} \\
\midrule    
    \multirow{5}{*}{$4.5 \pm 0.2$} & FTCP & \begin{tabular}{c}$0.00 $\\ $\pm0.00$\end{tabular} & \begin{tabular}{c}$0.00 $\\ $\pm0.00$\end{tabular} & \begin{tabular}{c}$\underline{\mathbf{1.00 }}$\\ $\underline{\mathbf{ \pm 0.00}}$\end{tabular} & \begin{tabular}{c}$0.20 $\\ $\pm0.04$\end{tabular} & \begin{tabular}{c}$0.99 $\\ $\pm0.00$\end{tabular} & \begin{tabular}{c}$0.20 $\\ $\pm0.04$\end{tabular} & \begin{tabular}{c}N/A\end{tabular} \\  \cdashline{2-9}
     & CDVAE & \begin{tabular}{c}$0.00 $\\ $\pm0.00$\end{tabular} & \begin{tabular}{c}$0.01 $\\ $\pm0.02$\end{tabular} & \begin{tabular}{c}$0.43 $\\ $\pm0.04$\end{tabular} & \begin{tabular}{c}$0.00 $\\ $\pm0.00$\end{tabular} & \begin{tabular}{c}$0.33 $\\ $\pm0.07$\end{tabular} & \begin{tabular}{c}$0.00 $\\ $\pm0.00$\end{tabular} & N/A \\  \cdashline{2-9}
     & TPE & \begin{tabular}{c}$0.02 $\\ $\pm0.01$\end{tabular} & \begin{tabular}{c}$\underline{\mathbf{0.86 }}$\\ $\underline{\mathbf{ \pm 0.04}}$\end{tabular} & \begin{tabular}{c}$0.35 $\\ $\pm0.03$\end{tabular} & \begin{tabular}{c}$0.21 $\\ $\pm0.01$\end{tabular} & \begin{tabular}{c}$0.31 $\\ $\pm0.02$\end{tabular} & \begin{tabular}{c}$\underline{\mathbf{1.00 }}$\\ $\underline{\mathbf{ \pm 0.00}}$\end{tabular} & \begin{tabular}{c}$1.00 $\\ $\pm0.00$\end{tabular} \\  \cdashline{2-9}
     & S(ALI) & \begin{tabular}{c}$\underline{\mathbf{0.36 }}$\\ $\underline{\mathbf{ \pm 0.03}}$\end{tabular} & \begin{tabular}{c}$0.59 $\\ $\pm0.03$\end{tabular} & \begin{tabular}{c}$1.00 $\\ $\pm0.00$\end{tabular} & \begin{tabular}{c}$\underline{\mathbf{0.50 }}$\\ $\underline{\mathbf{ \pm 0.03}}$\end{tabular} & \begin{tabular}{c}$\underline{\mathbf{1.00 }}$\\ $\underline{\mathbf{ \pm 0.00}}$\end{tabular} & \begin{tabular}{c}$\underline{\mathbf{1.00 }}$\\ $\underline{\mathbf{ \pm 0.00}}$\end{tabular} & \begin{tabular}{c}$0.17 $\\ $\pm0.03$\end{tabular} \\  \cdashline{2-9}
     & S(Cry) & \begin{tabular}{c}$0.26 $\\ $\pm0.04$\end{tabular} & \begin{tabular}{c}$0.55 $\\ $\pm0.02$\end{tabular} & \begin{tabular}{c}$0.97 $\\ $\pm0.02$\end{tabular} & \begin{tabular}{c}$0.48 $\\ $\pm0.06$\end{tabular} & \begin{tabular}{c}$\underline{\mathbf{1.00 }}$\\ $\underline{\mathbf{ \pm 0.00}}$\end{tabular} & \begin{tabular}{c}$\underline{\mathbf{1.00 }}$\\ $\underline{\mathbf{ \pm 0.00}}$\end{tabular} & \begin{tabular}{c}$0.68 $\\ $\pm0.01$\end{tabular} \\
\midrule    
    \multirow{5}{*}{$5.0 \pm 0.2$} & FTCP & \begin{tabular}{c}$0.00 $\\ $\pm0.00$\end{tabular} & \begin{tabular}{c}$0.00 $\\ $\pm0.01$\end{tabular} & \begin{tabular}{c}$\underline{\mathbf{1.00 }}$\\ $\underline{\mathbf{ \pm 0.00}}$\end{tabular} & \begin{tabular}{c}$0.21 $\\ $\pm0.01$\end{tabular} & \begin{tabular}{c}$0.98 $\\ $\pm0.01$\end{tabular} & \begin{tabular}{c}$0.21 $\\ $\pm0.01$\end{tabular} & \begin{tabular}{c}N/A\end{tabular} \\  \cdashline{2-9}
     & CDVAE & \begin{tabular}{c}$0.00 $\\ $\pm0.00$\end{tabular} & \begin{tabular}{c}$0.00 $\\ $\pm0.00$\end{tabular} & \begin{tabular}{c}$0.41 $\\ $\pm0.04$\end{tabular} & \begin{tabular}{c}$0.00 $\\ $\pm0.00$\end{tabular} & \begin{tabular}{c}$0.31 $\\ $\pm0.07$\end{tabular} & \begin{tabular}{c}$0.00 $\\ $\pm0.00$\end{tabular} & N/A \\  \cdashline{2-9}
     & TPE & \begin{tabular}{c}$0.03 $\\ $\pm0.01$\end{tabular} & \begin{tabular}{c}$0.65 $\\ $\pm0.07$\end{tabular} & \begin{tabular}{c}$0.35 $\\ $\pm0.04$\end{tabular} & \begin{tabular}{c}$0.20 $\\ $\pm0.04$\end{tabular} & \begin{tabular}{c}$0.30 $\\ $\pm0.03$\end{tabular} & \begin{tabular}{c}$\underline{\mathbf{1.00 }}$\\ $\underline{\mathbf{ \pm 0.00}}$\end{tabular} & \begin{tabular}{c}$0.92 $\\ $\pm0.17$\end{tabular} \\  \cdashline{2-9}
     & S(ALI) & \begin{tabular}{c}$\underline{\mathbf{0.30 }}$\\ $\underline{\mathbf{ \pm 0.02}}$\end{tabular} & \begin{tabular}{c}$\underline{\mathbf{0.80 }}$\\ $\underline{\mathbf{ \pm 0.01}}$\end{tabular} & \begin{tabular}{c}$1.00 $\\ $\pm0.00$\end{tabular} & \begin{tabular}{c}$0.40 $\\ $\pm0.04$\end{tabular} & \begin{tabular}{c}$\underline{\mathbf{1.00 }}$\\ $\underline{\mathbf{ \pm 0.00}}$\end{tabular} & \begin{tabular}{c}$\underline{\mathbf{1.00 }}$\\ $\underline{\mathbf{ \pm 0.00}}$\end{tabular} & \begin{tabular}{c}$0.24 $\\ $\pm0.01$\end{tabular} \\  \cdashline{2-9}
     & S(Cry) & \begin{tabular}{c}$0.26 $\\ $\pm0.00$\end{tabular} & \begin{tabular}{c}$0.60 $\\ $\pm0.03$\end{tabular} & \begin{tabular}{c}$0.96 $\\ $\pm0.01$\end{tabular} & \begin{tabular}{c}$\underline{\mathbf{0.45 }}$\\ $\underline{\mathbf{ \pm 0.03}}$\end{tabular} & \begin{tabular}{c}$\underline{\mathbf{1.00 }}$\\ $\underline{\mathbf{ \pm 0.00}}$\end{tabular} & \begin{tabular}{c}$\underline{\mathbf{1.00 }}$\\ $\underline{\mathbf{ \pm 0.00}}$\end{tabular} & \begin{tabular}{c}$0.61 $\\ $\pm0.03$\end{tabular} \\
\bottomrule
\end{tabular}
\end{table}

\clearpage

\subsection{Details in scaling to large atomic configurations: optimization of 135-atom perovskites} \label{appendix__details_of_large_system}

We conducted experiments on $3 \times 3 \times 3$ perovskite structures containing 135 atomic sites, expanded from a unit cell with five atomic sites. Because the cell size was larger, the range for the crystal lattice dimensions $a, b, c$ in SMOACS and TPE was set to 6--30~\r{A} for the $3 \times 3 \times 3$ structure. Similarly, the range of coordinate variations $\epsilon$ was set to $0.05$. All methods used perovskite structures with typical atomic coordinates and randomly generated lattice constants $a$, $b$, and $c$ in the range of 6~\r{A} to 30~\r{A}. Aside from these changes, the experimental conditions remained consistent with those described in Section~\ref{sec__perov_experiments}.

\begin{table}[hb]
\centering
\caption{Results for optimization of various band gaps while preserving a $3 \times 3 \times 3$ perovskite structure. Evaluation methods are based on those described in Table~\ref{table__perovskite_eval}.}
\label{table__3x3x3_supercell_all}
\begin{tabular}{llwc{1.3cm}wc{1.0cm}wc{1.0cm}wc{1.0cm}wc{1.0cm}wc{1.0cm}wc{1.0cm}wc{1.0cm}wc{1.0cm}}
\toprule
    \begin{tabular}{c}target $E_g$ (eV) \\ (Percentile) \end{tabular} & Method & SC rate &(i)$E_g$ &(ii)$E_f$ &  (iii) $t$ &(iv)Neu. & (v)Prv. \\
   \midrule \multirow{5}{*} {\begin{tabular}{c}$0.5 \pm 0.2$\\(47.4\%)\end{tabular}} & FTCP & \begin{tabular}{c}$0.00 $\\ $\pm0.00$\end{tabular} & \begin{tabular}{c}$0.84 $\\ $\pm0.02$\end{tabular} & \begin{tabular}{c}$\underline{\mathbf{1.00 }}$\\ $\underline{\mathbf{ \pm 0.00}}$\end{tabular} & \begin{tabular}{c}$0.00 $\\ $\pm0.00$\end{tabular} & N/A & \begin{tabular}{c}$0.00 $\\ $\pm0.00$\end{tabular} \\  \cdashline{2-8}
     & CDVAE & \begin{tabular}{c}$0.00 $\\ $\pm0.00$\end{tabular} & \begin{tabular}{c}$0.06 $\\ $\pm0.01$\end{tabular} & \begin{tabular}{c}$0.20 $\\ $\pm0.01$\end{tabular} & \begin{tabular}{c}$0.00 $\\ $\pm0.00$\end{tabular} & N/A & \begin{tabular}{c}$0.00 $\\ $\pm0.00$\end{tabular} \\  \cdashline{2-8}
     & TPE & \begin{tabular}{c}$0.00 $\\ $\pm0.00$\end{tabular} & \begin{tabular}{c}$\underline{\mathbf{1.00 }}$\\ $\underline{\mathbf{ \pm 0.00}}$\end{tabular} & \begin{tabular}{c}$0.00 $\\ $\pm0.00$\end{tabular} & \begin{tabular}{c}$\underline{\mathbf{0.99 }}$\\ $\underline{\mathbf{ \pm 0.02}}$\end{tabular} & N/A & \begin{tabular}{c}$\underline{\mathbf{1.00 }}$\\ $\underline{\mathbf{ \pm 0.00}}$\end{tabular} \\  \cdashline{2-8}
     & S(ALI) & \begin{tabular}{c}$\underline{\mathbf{0.36 }}$\\ $\underline{\mathbf{ \pm 0.00}}$\end{tabular} & \begin{tabular}{c}$0.92 $\\ $\pm0.01$\end{tabular} & \begin{tabular}{c}$1.00 $\\ $\pm0.01$\end{tabular} & \begin{tabular}{c}$0.40 $\\ $\pm0.00$\end{tabular} & \begin{tabular}{c}$\underline{\mathbf{1.00 }}$\\ $\underline{\mathbf{ \pm 0.00}}$\end{tabular} & \begin{tabular}{c}$\underline{\mathbf{1.00 }}$\\ $\underline{\mathbf{ \pm 0.00}}$\end{tabular} \\  \cdashline{2-8}
     & S(Cry) & \begin{tabular}{c}$0.05 $\\ $\pm0.03$\end{tabular} & \begin{tabular}{c}$0.93 $\\ $\pm0.03$\end{tabular} & \begin{tabular}{c}$0.10 $\\ $\pm0.04$\end{tabular} & \begin{tabular}{c}$0.39 $\\ $\pm0.03$\end{tabular} & \begin{tabular}{c}$\underline{\mathbf{1.00 }}$\\ $\underline{\mathbf{ \pm 0.00}}$\end{tabular} & \begin{tabular}{c}$\underline{\mathbf{1.00 }}$\\ $\underline{\mathbf{ \pm 0.00}}$\end{tabular} \\
\midrule    
    \multirow{5}{*} {\begin{tabular}{c}$4.0 \pm 0.2$\\(91.6\%)\end{tabular}} & FTCP & \begin{tabular}{c}$0.00 $\\ $\pm0.00$\end{tabular} & \begin{tabular}{c}$0.00 $\\ $\pm0.00$\end{tabular} & \begin{tabular}{c}$\underline{\mathbf{1.00 }}$\\ $\underline{\mathbf{ \pm 0.00}}$\end{tabular} & \begin{tabular}{c}$0.00 $\\ $\pm0.00$\end{tabular} & N/A & \begin{tabular}{c}$0.00 $\\ $\pm0.00$\end{tabular} \\  \cdashline{2-8}
     & CDVAE & \begin{tabular}{c}$0.00 $\\ $\pm0.00$\end{tabular} & \begin{tabular}{c}$0.00 $\\ $\pm0.00$\end{tabular} & \begin{tabular}{c}$0.24 $\\ $\pm0.02$\end{tabular} & \begin{tabular}{c}$0.00 $\\ $\pm0.00$\end{tabular} & N/A & \begin{tabular}{c}$0.00 $\\ $\pm0.00$\end{tabular} \\  \cdashline{2-8}
     & TPE & \begin{tabular}{c}$0.00 $\\ $\pm0.00$\end{tabular} & \begin{tabular}{c}$0.00 $\\ $\pm0.00$\end{tabular} & \begin{tabular}{c}$0.00 $\\ $\pm0.00$\end{tabular} & \begin{tabular}{c}$0.22 $\\ $\pm0.03$\end{tabular} & N/A & \begin{tabular}{c}$\underline{\mathbf{1.00 }}$\\ $\underline{\mathbf{ \pm 0.00}}$\end{tabular} \\  \cdashline{2-8}
     & S(ALI) & \begin{tabular}{c}$\underline{\mathbf{0.42 }}$\\ $\underline{\mathbf{ \pm 0.01}}$\end{tabular} & \begin{tabular}{c}$\underline{\mathbf{0.75 }}$\\ $\underline{\mathbf{ \pm 0.01}}$\end{tabular} & \begin{tabular}{c}$\underline{\mathbf{1.00 }}$\\ $\underline{\mathbf{ \pm 0.00}}$\end{tabular} & \begin{tabular}{c}$\underline{\mathbf{0.47 }}$\\ $\underline{\mathbf{ \pm 0.01}}$\end{tabular} & \begin{tabular}{c}$\underline{\mathbf{1.00 }}$\\ $\underline{\mathbf{ \pm 0.00}}$\end{tabular} & \begin{tabular}{c}$\underline{\mathbf{1.00 }}$\\ $\underline{\mathbf{ \pm 0.00}}$\end{tabular} \\  \cdashline{2-8}
     & S(Cry) & \begin{tabular}{c}$0.03 $\\ $\pm0.02$\end{tabular} & \begin{tabular}{c}$0.17 $\\ $\pm0.04$\end{tabular} & \begin{tabular}{c}$0.11 $\\ $\pm0.02$\end{tabular} & \begin{tabular}{c}$0.44 $\\ $\pm0.06$\end{tabular} & \begin{tabular}{c}$\underline{\mathbf{1.00 }}$\\ $\underline{\mathbf{ \pm 0.00}}$\end{tabular} & \begin{tabular}{c}$\underline{\mathbf{1.00 }}$\\ $\underline{\mathbf{ \pm 0.00}}$\end{tabular} \\
\midrule    
    \multirow{5}{*} {\begin{tabular}{c}$4.5 \pm 0.2$\\(94.4\%)\end{tabular}} & FTCP & \begin{tabular}{c}$0.00 $\\ $\pm0.00$\end{tabular} & \begin{tabular}{c}$0.00 $\\ $\pm0.00$\end{tabular} & \begin{tabular}{c}$\underline{\mathbf{1.00 }}$\\ $\underline{\mathbf{ \pm 0.00}}$\end{tabular} & \begin{tabular}{c}$0.00 $\\ $\pm0.00$\end{tabular} & N/A & \begin{tabular}{c}$0.00 $\\ $\pm0.00$\end{tabular} \\  \cdashline{2-8}
     & CDVAE & \begin{tabular}{c}$0.00 $\\ $\pm0.00$\end{tabular} & \begin{tabular}{c}$0.00 $\\ $\pm0.00$\end{tabular} & \begin{tabular}{c}$0.21 $\\ $\pm0.04$\end{tabular} & \begin{tabular}{c}$0.00 $\\ $\pm0.00$\end{tabular} & N/A & \begin{tabular}{c}$0.00 $\\ $\pm0.00$\end{tabular} \\  \cdashline{2-8}
     & TPE & \begin{tabular}{c}$0.00 $\\ $\pm0.00$\end{tabular} & \begin{tabular}{c}$0.00 $\\ $\pm0.00$\end{tabular} & \begin{tabular}{c}$0.00 $\\ $\pm0.00$\end{tabular} & \begin{tabular}{c}$0.21 $\\ $\pm0.04$\end{tabular} & N/A & \begin{tabular}{c}$\underline{\mathbf{1.00 }}$\\ $\underline{\mathbf{ \pm 0.00}}$\end{tabular} \\  \cdashline{2-8}
     & S(ALI) & \begin{tabular}{c}$\underline{\mathbf{0.33 }}$\\ $\underline{\mathbf{ \pm 0.03}}$\end{tabular} & \begin{tabular}{c}$\underline{\mathbf{0.76 }}$\\ $\underline{\mathbf{ \pm 0.04}}$\end{tabular} & \begin{tabular}{c}$1.00 $\\ $\pm0.01$\end{tabular} & \begin{tabular}{c}$0.42 $\\ $\pm0.02$\end{tabular} & \begin{tabular}{c}$\underline{\mathbf{1.00 }}$\\ $\underline{\mathbf{ \pm 0.00}}$\end{tabular} & \begin{tabular}{c}$\underline{\mathbf{1.00 }}$\\ $\underline{\mathbf{ \pm 0.00}}$\end{tabular} \\  \cdashline{2-8}
     & S(Cry) & \begin{tabular}{c}$0.06 $\\ $\pm0.02$\end{tabular} & \begin{tabular}{c}$0.21 $\\ $\pm0.03$\end{tabular} & \begin{tabular}{c}$0.14 $\\ $\pm0.03$\end{tabular} & \begin{tabular}{c}$\underline{\mathbf{0.44 }}$\\ $\underline{\mathbf{ \pm 0.05}}$\end{tabular} & \begin{tabular}{c}$\underline{\mathbf{1.00 }}$\\ $\underline{\mathbf{ \pm 0.00}}$\end{tabular} & \begin{tabular}{c}$\underline{\mathbf{1.00 }}$\\ $\underline{\mathbf{ \pm 0.00}}$\end{tabular} \\
\midrule    
    \multirow{5}{*} {\begin{tabular}{c}$5.0 \pm 0.2$\\(96.5\%)\end{tabular}} & FTCP & \begin{tabular}{c}$0.00 $\\ $\pm0.00$\end{tabular} & \begin{tabular}{c}$0.00 $\\ $\pm0.00$\end{tabular} & \begin{tabular}{c}$\underline{\mathbf{1.00 }}$\\ $\underline{\mathbf{ \pm 0.00}}$\end{tabular} & \begin{tabular}{c}$0.00 $\\ $\pm0.00$\end{tabular} & N/A & \begin{tabular}{c}$0.00 $\\ $\pm0.00$\end{tabular} \\  \cdashline{2-8}
     & CDVAE & \begin{tabular}{c}$0.00 $\\ $\pm0.00$\end{tabular} & \begin{tabular}{c}$0.00 $\\ $\pm0.00$\end{tabular} & \begin{tabular}{c}$0.24 $\\ $\pm0.01$\end{tabular} & \begin{tabular}{c}$0.00 $\\ $\pm0.00$\end{tabular} & N/A & \begin{tabular}{c}$0.00 $\\ $\pm0.00$\end{tabular} \\  \cdashline{2-8}
     & TPE & \begin{tabular}{c}$0.00 $\\ $\pm0.00$\end{tabular} & \begin{tabular}{c}$0.00 $\\ $\pm0.00$\end{tabular} & \begin{tabular}{c}$0.00 $\\ $\pm0.00$\end{tabular} & \begin{tabular}{c}$0.18 $\\ $\pm0.03$\end{tabular} & N/A & \begin{tabular}{c}$\underline{\mathbf{1.00 }}$\\ $\underline{\mathbf{ \pm 0.00}}$\end{tabular} \\  \cdashline{2-8}
     & S(ALI) & \begin{tabular}{c}$\underline{\mathbf{0.35 }}$\\ $\underline{\mathbf{ \pm 0.01}}$\end{tabular} & \begin{tabular}{c}$\underline{\mathbf{0.79 }}$\\ $\underline{\mathbf{ \pm 0.04}}$\end{tabular} & \begin{tabular}{c}$\underline{\mathbf{1.00 }}$\\ $\underline{\mathbf{ \pm 0.00}}$\end{tabular} & \begin{tabular}{c}$\underline{\mathbf{0.41 }}$\\ $\underline{\mathbf{ \pm 0.00}}$\end{tabular} & \begin{tabular}{c}$\underline{\mathbf{1.00 }}$\\ $\underline{\mathbf{ \pm 0.00}}$\end{tabular} & \begin{tabular}{c}$\underline{\mathbf{1.00 }}$\\ $\underline{\mathbf{ \pm 0.00}}$\end{tabular} \\  \cdashline{2-8}
     & S(Cry) & \begin{tabular}{c}$0.04 $\\ $\pm0.02$\end{tabular} & \begin{tabular}{c}$0.15 $\\ $\pm0.05$\end{tabular} & \begin{tabular}{c}$0.12 $\\ $\pm0.02$\end{tabular} & \begin{tabular}{c}$0.41 $\\ $\pm0.03$\end{tabular} & \begin{tabular}{c}$\underline{\mathbf{1.00 }}$\\ $\underline{\mathbf{ \pm 0.00}}$\end{tabular} & \begin{tabular}{c}$\underline{\mathbf{1.00 }}$\\ $\underline{\mathbf{ \pm 0.00}}$\end{tabular} \\
\bottomrule
\end{tabular}
\end{table}

\clearpage

\subsection{Cross-dataset multi-objective constrained optimization with SMOACS} \label{section__cross_dataset}

A key strength of SMOACS is its ability to straightforwardly combine prediction models trained on different datasets without retraining. This flexibility allows multi-objective optimization by simply swapping in pretrained models---such as those available for ALIGNN. We optimized perovskite structures using two settings: (i) a MEGNet-based formation-energy predictor with a bulk-modulus predictor trained on the JARVIS-DFT dataset, and (ii) a MEGNet band-gap predictor with an $E_\mathrm{hull}$ predictor trained on the JARVIS-DFT dataset.

\begin{table}[hp]
\centering
\caption{Results of optimizing perovskite structures for targeted band gaps using a MEGNet band gap predictor and a JARVIS-DFT $E_{\mathrm{hull}}$ predictor. Success rate is defined as satisfying all five criteria (i)-(v). Criteria (i), (iii)-(v) follow Table~\ref{table__perovskite_eval}; (ii) requires $E_{\mathrm{hull}} < 0.1$ eV/atom.}
\label{table__smoacs_ali_megnet-bg_jarvis_e_full}
\begin{tabular}{lcccccc}
\toprule
     \begin{tabular}{c}Target $E_g$ (Percentile)\end{tabular}& Success rate & (i)$E_g$ & (ii)$E_\mathrm{hull}$ & (iii) $t$ & (iv)Neu. &  (v)Prv. \\  \midrule 
      \begin{tabular}{c}$4.0 \pm 0.2$ eV (91.6\%)\end{tabular} & $0.52\pm0.05$ & 0.60 & 0.87 & 0.82 & 1.00 & 1.00 \\
    \begin{tabular}{c}$4.5 \pm 0.2$ eV (94.4\%)\end{tabular} & $0.54\pm0.02$ & 0.62 & 0.86 & 0.82 & 1.00 & 1.00 \\
    \begin{tabular}{c}$5.0 \pm 0.2$ eV (96.5\%)\end{tabular} & $0.51\pm0.03$ & 0.60 & 0.84 & 0.79 & 1.00 & 1.00 \\
\bottomrule
\end{tabular}
\end{table}

\begin{table}[ht]
\centering
\caption{Results of optimizing perovskite structures for targeted bulk modulus (denoted as "$B$") values using a JARVIS-DFT based bulk modulus predictor and a MEGNet-based formation energy predictor. Success rate is defined as satisfying all five criteria (i)-(v). Criteria (ii)-(v) follow Table~\ref{table__perovskite_eval}.}
\label{table__smoacs_ali_megnet-Ef_jarvis_BM}
\begin{tabular}{lcccccc}
\midrule
  \begin{tabular}{c}Target $B$ (Percentile)\end{tabular}& Success rate & (i) $B$ & (ii)$E_f$ & (iii) $t$ & (iv)Neu. &  (v)Prv. \\  \midrule 
      \begin{tabular}{c}$120.0 \pm 10.0$ GPa (75.8\%)\end{tabular} & $0.31\pm0.05$ & 0.68 & 0.80 & 0.55 & 1.00 & 1.00 \\
    \begin{tabular}{c}$160.0 \pm 10.0$ GPa (86.6\%)\end{tabular} & $0.34\pm0.04$ & 0.77 & 0.70 & 0.65 & 1.00 & 1.00 \\
    \begin{tabular}{c}$200.0 \pm 10.0$ GPa (93.7\%)\end{tabular} & $0.10\pm0.02$ & 0.76 & 0.41 & 0.27 & 1.00 & 1.00 \\
\bottomrule
\end{tabular}
\end{table}

As shown in Tables~\ref{table__smoacs_ali_megnet-bg_jarvis_e_full} and~\ref{table__smoacs_ali_megnet-Ef_jarvis_BM}, SMOACS successfully proposed perovskite structures that satisfy all targets and constraints, even when combining predictors trained on different datasets. This demonstrates its robustness in cross-dataset multi-objective optimization. 

\clearpage

\subsection{Space-group conditional generation benchmark against MatterGen}
\label{appendix___spg_conditioned_generation}

SMOACS performs space-group-conditional generation by optimizing only symmetry-allowed variables. The lattice was initialized in a lattice system compatible with the target space group, and only the independent lattice parameters were updated. Atomic sites are initialized at Wyckoff positions and updated only through their free Wyckoff parameters. The full structure is reconstructed from space-group operations during optimization. For this benchmark, SMOACS minimized energy using a formation-energy predictor trained on the MEGNet dataset. The optimized candidates were then relaxed with MatterSim, and their SPGs were assigned by pymatgen. S.U.N. was evaluated using the same protocol as in Section~\ref{sec__comparison_with_mattergen}.

\begin{table}[!b]
\centering
\caption{Comparison of SMOACS and MatterGen under space-group-conditional generation. The SPG ratio is the fraction of structures in the MatterGen pretraining data that have the target space group (SPG). Each result was obtained from 128 samples for each target SPG. ``SPG \& S.U.N.'' denotes the fraction of samples that simultaneously satisfy both the target SPG and the Stability--Uniqueness--Novelty criteria. ``SPG SC rate'' denotes the target-SPG success rate, and ``S.U.N. SC rate'' denotes the S.U.N. success rate.}
\label{table__mattergen_spg}
\begin{tabular}{wl{2.7cm}wc{1.1cm}wl{1.25cm}wc{1.2cm}wc{1.2cm}wc{1.2cm}}
\toprule
\makecell{Target\\space group}
& \makecell{MatterGen\\pretrain\\SPG ratio}
& Method
& \makecell{SPG\&\\S.U.N.}
& \makecell{SPG\\SC rate}
& \makecell{S.U.N.\\SC rate} \\
\midrule
\multirow{2}{*}{\makecell[l]{$\mathrm{P}321$ (No.~150)\\trigonal}}
& \multirow{2}{*}{0.000}
& MatterGen & 0.000 & 0.000 & 0.605 \\
& & SMOACS    & \underline{\textbf{0.031}} & 0.547 & 0.055 \\
\midrule
\multirow{2}{*}{\makecell[l]{$\mathrm{P}6_{3}/\mathrm{m}$ (No.~176)\\hexagonal}}
& \multirow{2}{*}{0.002}
& MatterGen & 0.002 & 0.012 & 0.615 \\
& & SMOACS    & \underline{\textbf{0.062}} & 0.898 & 0.070 \\
\midrule
\multirow{2}{*}{\makecell[l]{$\mathrm{Amm}2$ (No.~38)\\orthorhombic}}
& \multirow{2}{*}{0.011}
& MatterGen & \underline{\textbf{0.236}} & 0.312 & 0.635 \\
& & SMOACS    & 0.039 & 0.953 & 0.047 \\
\midrule
\multirow{2}{*}{\makecell[l]{$\mathrm{R}3\mathrm{m}$ (No.~160)\\trigonal}}
& \multirow{2}{*}{0.018}
& MatterGen & \underline{\textbf{0.316}} & 0.410 & 0.684 \\
& & SMOACS    & 0.070 & 0.805 & 0.070 \\
\midrule
\multirow{2}{*}{\makecell[l]{$\mathrm{P}2/\mathrm{m}$ (No.~10)\\monoclinic}}
& \multirow{2}{*}{0.036}
& MatterGen & \underline{\textbf{0.307}} & 0.361 & 0.768 \\
& & SMOACS    & 0.055 & 0.742 & 0.094 \\
\midrule
\multirow{2}{*}{\makecell[l]{$\mathrm{P}4/\mathrm{mmm}$ (No.~123)\\tetragonal}}
& \multirow{2}{*}{0.092}
& MatterGen & \underline{\textbf{0.574}} & 0.674 & 0.699 \\
& & SMOACS    & 0.078 & 0.797 & 0.078 \\
\bottomrule
\end{tabular}
\end{table}

Table~\ref{table__mattergen_spg} shows that MatterGen maintains high S.U.N. SC rates, but its space-group satisfaction rate (SPG SC rate) tends to decrease for rarer target space groups in its pretraining data. SMOACS can impose the target SPG during generation even when no crystal with that SPG appears in the predictor training data, because the SPG is imposed through the optimization domain rather than learned from data. Its lower joint scores are mainly limited by S.U.N.

Although the listed SMOACS SPG SC rates are below 1.0, SMOACS achieves 100\% space-group preservation in principle. The structures counted as space-group failures were confirmed to relax into higher-symmetry space groups that contain the target symmetry. Thus, the reported SPG SC rate measures exact post-relaxation space-group labels assigned by pymatgen, not loss of the imposed symmetry.

\clearpage

\subsection{Band-gap targeting benchmark without \(L_{\mathrm{elm}}\)}
\label{appendix__mattergen_bandgap_without_l_elm}

For the ablation study in this subsection, \(L_{\mathrm{elm}}\) was removed from the S.U.N.-BG objective. Specifically, the optimization was performed using the following reduced objective:
\begin{align}
\mathcal{L}'_{\mathrm{S.U.N.-BG}} = L_g + L_f.
\label{eq__methods_matgen_bg_compare_wo_l_elm}
\end{align}

\begin{table}[hp]
\centering
\caption{DFT-assessed band-gap targeting benchmark without \(L_{\mathrm{elm}}\). The notations follow Table~\ref{table__mattergen_bandgap}. For each target and method, we evaluated 128 samples.}
\label{table__mattergen_bandgap_without_l_elm}
\begin{tabular}{wl{0.85cm}wl{1.2cm}wc{1.2cm}wc{1.0cm}wc{1.0cm}}
\toprule
\makecell{Target\\$E_g$ (eV)} & Method & \makecell{BG\& \\ S.U.N.} & \makecell{BG\\SC rate} & \makecell{S.U.N.\\SC rate} \\
\midrule
\multirow{2}{*}{\makecell{$2.5$\\$\pm0.2$}}
& MatterGen & \underline{\textbf{0.039}} & 0.055 & 0.680 \\
& SMOACS    & 0.000 & 0.023 & 0.117 \\
\midrule
\multirow{2}{*}{\makecell{$4.0$\\$\pm0.2$}}
& MatterGen & 0.023 & 0.039 & 0.562 \\
& SMOACS    & \underline{\textbf{0.047}} & 0.078 & 0.406 \\
\midrule
\multirow{2}{*}{\makecell{$4.5$\\$\pm0.2$}}
& MatterGen & 0.023 & 0.039 & 0.516 \\
& SMOACS    & \underline{\textbf{0.047}} & 0.109 & 0.336 \\
\midrule
\multirow{2}{*}{\makecell{$5.0$\\$\pm0.2$}}
& MatterGen & 0.047 & 0.070 & 0.547 \\
& SMOACS    & 0.047 & 0.062 & 0.328 \\
\bottomrule
\end{tabular}
\end{table}

\end{appendices}

\end{document}